\documentclass[manuscript]{aastex}
\usepackage{pxfonts}
\usepackage{color}

\shorttitle{Emissions from Cygnus X--1}
\shortauthors{Zhang, Xu \& Lu}

\begin{document}


\title{Origin of Multi-band Emission from the Microquasar Cygnus X--1}


\author{Jianfu Zhang\altaffilmark{1,2}, Bing Xu\altaffilmark{2}, Jufu Lu\altaffilmark{1}}
%,
%%\affil{Astronomy Department, University of California,
%%    Berkeley, CA 94720}
%%\author{C. D. Biemesderfer\altaffilmark{4,5}}
%%\affil{National Optical Astronomy Observatories, Tucson, AZ 85719}
%%\email{lizhang@ynu.edu.cn}
%%\and
%%\author{R. J. Hanisch\altaffilmark{5}}
%%\affil{Space Telescope Science Institute, Baltimore, MD 21218}


\altaffiltext{1}{Department of Astronomy and Institute of Theoretical Physics and Astrophysics, Xiamen University, Xiamen, Fujian 361005, China;}
 \email{jianfuzhang.yn@gmail.com}  %lujf@xmu.edu.cn
\altaffiltext{2}{Department of Physics, Tongren University, Tongren 554300, China}
%%\altaffiltext{4}{Visiting Programmer, Space Telescope Science Institute}
%%\altaffiltext{5}{Patron, Alonso's Bar and Grill}


\begin{abstract}
We study the origin of non-thermal emissions from the Galactic black hole X-ray binary Cygnus X--1 which is a confirmed high mass microquasar. By analogy with those methods used in studies of active galactic nuclei, we propose a two-dimensional, time-dependent radiation model from the microquasar Cygnus X--1. In this model, the evolution equation for relativistic electrons in a conical jet are numerically solved by including escape, adiabatic and various radiative losses. The radiative processes involved are synchrotron emission, its self-Compton scattering, and inverse Compton scatterings of an accretion disk and its surrounding stellar companion. This model also includes an electromagnetic cascade process of an anisotropic $\gamma$-$\gamma$ interaction. We study the spectral properties of electron evolution and its emission spectral characteristic at different heights of the emission region located in the jet. We find that radio data from Cygnus X--1 are reproduced by the synchrotron emission, the \emph{Fermi} LAT measurements by the synchrotron emission and Comptonization of photons of the stellar companion, the TeV band emission fluxes by the Comptonization of the stellar photons. Our results show that:  (1) Radio emission region extends from the binary system scales to the termination of the jet. (2) The GeV band emissions should originate from the distance close to the binary system scales. (3) The TeV band emissions could be inside the binary system, and these emissions could be probed by the upcoming CTA telescope. (4) The MeV tail emissions, which produces a strongly linearly polarized signal, are emitted inside the binary system. The location of the emissions is very close to the inner region of the jet.
\end{abstract}



\section{INTRODUCTION}
Cygnus X--1 is a Galactic black hole X-ray binary which is composed of a black hole and a high mass stellar companion. According to the properties of radio jets, it was classified as the microquasar (\citealt{Mirabel99}) and has been confirmed. The studies of the spectral behavior at the X-ray bands reveal that Cygnus X--1 exhibits three typically spectral states. The soft state is characterized by a relatively bright spectrum, a strongly thermal radiation characteristic and a steep power-law tail, which can be modeled by the multi-temperature blackbody radiation of a standard thin accretion disk (\citealt{Shakura73}) in the soft-X-ray band. Cygnus X--1 spends most of the time in the hard spectral state whose spectral energy distribution is described by a power law plus a exponential cutoff at about $150\ \rm keV$. The intermediate state usually associates with the state transition of the above two states, and is a combination of both states (\citealt{Remillard06,Belloni10}).

In the hard spectral state, the broadband observations (radio to $\gamma$-ray bands) have been analyzed and collected in the literature (\citealt{ZAA12,ZAA14b}). Radio observations from the microquasar Cygnus X--1 support the presence of the persistent, relativistic jets. At present, the existence of leptons in jets are generally approved, but a little evidence on hadron components has only been provided (SS 433: \citealt{Migliari02}; 4U 1630¨C47: \citealt{Diaz13}, see also \citealt{Neilsen14} for a disputation). The theoretical studies on jets involve hadronic and/or leptonic contents. In the hadronic model, protons in jets are accelerated to very high energy to produce $\gamma$-ray photons (e.g.,\citealt{Romero03,Romero08,Zhang11}) and neutrinos (e.g., \citealt{Zhang10}) by the proton-proton and/or proton-photon collisions. In the leptonic model, the electrons accelerated in jets can emit the multiwavelength radiation by synchrotron and inverse Compton scattering processes (\citealt{Bosch06,Peer09,ZAA12,Malyshev13,ZAA14a}). Recently, more comprehensive comments regarding theoretical and observational progresses have been done in this field (e.g., \citealt{Bosch09,Dubus13,Bednarek13}).

In the framework of leptonic models, by employing numerical methods and assuming a steady state electron distribution, \cite{Bosch06} carried out a comprehensive study on spectral energy distributions produced in jets of microquasars. Based on this work, the anisotropic inverse Compton scattering effects have been considered in \cite{Zhang09}. Recently, a series of works based on analytical and numerical methods are performed in the black hole X-ray binaries (e.g., \citealt{ZAA12,Malyshev13,ZAA14a,ZAA14b}). In this paper, by analogy with those methods used in extragalactic blazar objects (\citealt{ms03}), we propose a two-dimensional, time-dependent radiation model from the microquasar Cygnus X--1 by solving the kinetic equation for relativistic electron. The spectral properties of evolution electrons are studied  by including the escape, adiabatic and various radiative coolings. The spectral distribution of photons, being produced at different locations of emission region in a jet, are presented. Furthermore, this model also includes electromagnetic cascade processes, and emission of pairs are calculated. To constrain a possible region of emissions in which what band radiation is produced, we carry out several fittings about Cygnus X--1 ranging from radio to high energy bands. According to these multiwavelength fittings, we can explore radiation mechanism of broadband emissions and limit their emission regions in the jet.

The structure of the paper is as follows. In the next section, we present the basic properties of the radiation model from Cygnus X--1; numerical results are given in Section 3; in Section 4, we give the spectral energy distributions by confronting with multiwavelength observations; section 5 contains the conclusion and discussions.

\section{Model Description}
\begin{figure}
\centerline{
\includegraphics[width=12.0cm]{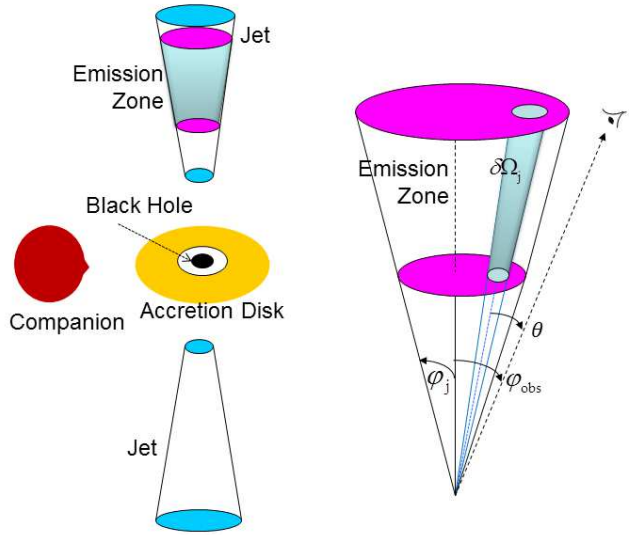}
}
\caption{Left panel: main components of the microquasar Cygnus X--1. Right panel: a schematic description of numerical calculation processes from an emission zone in the jet, indicating related geometrical parameters. The half-opening angle $\varphi_{\rm j}$ is divided into various angle-intervals with each thick $\delta \Omega_{\rm j}$ that contains $\delta N_{\rm \gamma}$ electrons. The symmetry axis of the jet make an angle $\varphi_{\rm obs}$ with the line of sight. The numerical calculations are carried out in the emission zone by summation over $\theta$ contribution.} \label{fig:topology}
\end{figure}

In this section, we present a radiation model from the microquasar Cygnus X--1 and numerical methods used in this work. We use a classical geometry of the microquasar in which dipolar jets are launched from the inner part of an accretion disk, perpendicularly to its orbital plane. In Figure \ref{fig:topology}, we present schematic diagrams of geometrical structure (left panel) and of numerical procedure (right panel) for Cygnus X--1.    Our numerical methods are similar to those of the extragalactic blazar object by \cite{ms03}. As will be seen, many significant modifications have been performed in order to match the scenarios in Cygnus X--1. For the convenience of the reader, we also present main radiation mechanism. Regarding relevant details, interested readers are referred to \cite{Aharonian81}, \cite{ CG99}, \cite{ms03} and \cite{ms05}.

\subsection{Electron Evolution Process}
Based on an assumption that energy densities of magnetic field and of external radiation field are homogenous across an emission region, and a consideration regarding a possible escape process of the accelerated electron, we regard the evolution equation for relativistic electron along the jets as
\begin{equation}
{\partial N_{\rm \gamma}(\gamma,z) \over \partial z} =-{\partial \over \partial
\gamma} \left[N_{\rm \gamma}(\gamma,z) {{\rm d}\gamma \over {\rm d}z}\right] + {Q
\over c \beta_{\Gamma} \Gamma}-{c \beta_{\Gamma}\Gamma
N_{\rm \gamma}(\gamma,z)\over z}\ . \label{dNdz}
\end{equation}
The last term on the right-hand side of Equation (\ref{dNdz}) is the escape loss rate of the accelerated electron. Where $\Gamma$ is the bulk Lorentz factor of the jet matter with $\beta_{\Gamma} = \sqrt{\Gamma^2-1}/\Gamma$. $\gamma$ is the electron Lorentz factor and $c$ is the speed of light. The energy change of the relativistic electron along the jet is given as
\begin{equation}
{{\rm d}\gamma \over {\rm d}z} = {1 \over  c\beta_{\Gamma} \Gamma}\left({\rm d}\gamma \over {\rm d}t^{'}\right)_{\rm rad}- {2 \over
3}{\gamma \over z}\ , \label{dgdz}
\end{equation}
where ${\rm d}t^{'}$ is the proper time. The second term of the right-hand side of Equation (\ref{dgdz}) stands for the adiabatic loss due to a two-dimensional expansion of the jet matter. The total radiative loss rate of an electron
\begin{equation}
{\left({\rm d}\gamma \over {\rm d}t^{'}\right)_{\rm rad}=\left({\rm
d}\gamma \over {\rm d}t^{'}\right)_{\rm Syn} +\left({\rm d}\gamma
\over {\rm d}t^{'}\right)_{\rm SSC}+\left({\rm d}\gamma \over {\rm
d}t^{'}\right)_{\rm EXT}} \label{dgdt}
\end{equation}
is to be discussed later in detail. In Equation (\ref{dNdz}), $Q$ is called injection function and is a characteristic for energy distributions of the accelerated electron. In this work, we assume that it follows a power-law distribution
\begin{equation}
Q (\gamma)=K\gamma^{-p},
\end{equation}
where $p$ is the power-law index of the accelerated electron; $K$ is the normalization constant which can be determined by
\begin{equation}
{L_{\rm rel}}=K\int^{\gamma_{\rm max}}_{\gamma_{\rm min}}\gamma^{-p}\gamma d\gamma. \label{Lrel}
\end{equation}
Where $\gamma_{\rm max}$ and $\gamma_{\rm min}$ are maximum and minimum energies of the relativistic electron, respectively. Here, we consider $\gamma_{\rm min}$ as a free parameter. In Equation (\ref{Lrel}), $L_{\rm rel}=q_{\rm rel}L_{\rm jet}$ is a fraction of power in the form of the relativistic electrons, in which $L_{\rm jet}=q_{\rm jet}L_{\rm acc}$ is fixed to be proportional to the accretion power $L_{\rm acc}=\dot{M}_{\rm acc} c^2$. Here, $q_{\rm jet}$ and $q_{\rm rel}$ are model free parameters, and $\dot{M}_{\rm acc}$ is the mass accretion rate.

\subsection{Synchrotron Emission}

The energy loss rate of an electron for synchrotron process is
\begin{equation}
\left({\rm d}\gamma \over {\rm d}t'\right)_{\rm Syn} = - {4c\sigma_{\rm
T} \over 3 m_{\rm e} c^2 } u_{\rm B}' \gamma^2\ ,
\label{dgdtsyn}
\end{equation}
where $\sigma_{\rm T}$, $m_{\rm e}$ are the Thomson cross section and rest mass of an electron, respectively. $u'_{B} = {B'^2\over 8\pi}$ is the energy density of magnetic field. Quantities measured in the jet frame are primed throughout the paper. According to conservation of magnetic flux in a jet, we have the the magnetic field strength of $B'= B_{0}^{'}({z_{\rm 0}\over z})$ along the jet, where $B_{\rm 0}^{'}$ is the magnetic field strength at the height of $z_{\rm 0}$ and is a free parameter in this work.

Concerning the integration procedure shown in Figure \ref{fig:topology}, this work is very similar to that of Figure 1 in \cite{ms03} for a two-dimensional scenario. An emission region is enclosed within the half-opening angle $\varphi_{\rm j}$ of the jet, corresponding to a solid angle $\Omega_{\rm j}$, which is divided into multiple angle-intervals with each thick of $\delta \Omega_{\rm j}$ that contains $\delta N_{\rm \gamma}$ electrons. Different than \cite{ms03} that is a one-zone model, the current model is flexible and can extend from an acceleration and emission region to anther extended emission region along the jet. The emissivity of synchrotron radiation in solid angle $\delta \Omega_{\rm j}$ for a given electron distribution is
(\citealt{CS86})
\begin{equation}
\delta J_{\nu'}'(z) = \frac{\sqrt{3}e^3}{4\pi m_{\rm e}c^2}B'(z)\int_{\gamma_{\rm min}}^{\gamma_{\rm
max}} \int_{\Omega}  \delta N_{\rm \gamma}(z) F_{\rm Syn}(\nu'/\nu'_{\rm c}) {\rm sin}\alpha {\rm d} \gamma {\rm d}\Omega\ .
 \label{LS}
\end{equation}
where $\alpha$ is the pitch angle of magnetic field line. The electron number in each solid angle is $\delta N_{\rm \gamma} = N_{\rm \gamma} (\delta \Omega_{\rm j}/\Omega_{\rm j})$ and $F_{\rm Syn}(\nu'/\nu'_{\rm c})=F_{\rm Syn}(x)=x\int_{x}^{\infty}K_{\rm 5/3}(y){\rm d}y$, in which $K_{\rm 5/3}$ is the Bessel function of the second kind. The characteristic frequency $\nu'_{\rm c}$ is expressed as $\nu'_{\rm c}=3eB'(z)\gamma^2 {\rm sin}\alpha/4\pi m_{\rm e}c$.
The synchrotron absorption coefficient in solid angle $\delta \Omega_{\rm j}$ is determined by
\begin{equation}
\delta K_{\nu'}'(z) = - \frac{\sqrt{3}e^3}{4\pi m_{\rm e }c^2}\frac{c^2}{2m_{\rm e}c^2\nu^{'2}}B'(z)\int_{\gamma_{\rm min}}^{\gamma_{\rm
max}} \int_{\Omega} \frac{\rm d}{\rm d \gamma}\left(\frac{\delta N_{\rm \gamma}(z)}{\gamma^2}\right ) \gamma^2F_{\rm Syn}(\nu'/\nu'_{\rm c}) {\rm sin}\alpha {\rm d} \gamma {\rm d}\Omega. \label{Absorption}
 \label{LS}
\end{equation}
The luminosity of synchrotron emission per frequency in solid angle $\delta \Omega_{\rm j}$ is written as
\begin{equation}
\delta L_{{\rm Syn},\nu'}'(z) = \frac{\delta J_{\nu'}'(z)}{\delta K_{\nu'}'(z)}[1-{\rm exp}(-\tau_{\rm self})]\ ,
\end{equation}
where $\tau_{\rm self}$ is the optical depth of synchrotron self absorption, which is determined by Equation (\ref{Absorption}).

The energy density of synchrotron photons in solid angle $\delta \Omega_{\rm j}$ being produced at each height $z$ of the emitting region can be approximated as (\citealt{ms03}
\begin{equation}
{\rm d} u_{\rm Syn}' \simeq {1\over 2  c z^{2} \delta \Omega_{\rm j}}\int \, \delta L_{{\rm Syn},\nu'}'(z) {\rm d}
\nu' \ .
\end{equation}
The total energy density of synchrotron emission at a certain height of the jet is the sum of itself and other distances before it, which is given by
\begin{equation}
u_{\rm Syn}'(z)= \int_{z_{\rm 0}}^{z} \left(\frac{z_{\rm 0}}{z}\right )^2  {\rm d} u_{\rm Syn}'\ .\label{SYND}
\end{equation}

\subsection{Inverse Compton Scattering}

Below, we are to depict that relativistic electrons Compton-upscatter seed photons in the Klein-Nishina regime. In the current scenario, three different seed photon fields are involved: synchrotron radiation field in a jet, a photon field of the stellar companion and a photon field of the accretion disk. The photon field of synchrotron radiation is considered as an anisotropic distribution in the jet. Moreover, due to the fact that the microquasar Cygnus X--1 has a particular geometry effect, the soft seed photons from the stellar companion and accretion disk are also strongly anisotropic. Thus we take into account an angle-dependent  cross-section in the Klein-Nishina regime (\citealt{Aharonian81,ms05}).

The energy loss rate of relativistic electron due to the synchrotron self-Compton (SSC) process and external inverse Compton process is written as
\begin{equation}
\left({\rm d} \gamma \over {\rm d} t'\right)_{i} = - {4
c\sigma_{T} \over 3 m_{\rm e} c^2 }\gamma^2 \int f_{\rm KN}(b)\varepsilon_{i} u'_{i}{\rm d} \varepsilon_{i} ,
\end{equation}
where the subscript $i$ (=Syn/SSC, star and disk) stands for the components from synchrotron emission, companion star and accretion disk, respectively.  $\varepsilon_{i}= h\nu'_{i}/m_{\rm e} c^2$ is the energy of a seed photon. The dimensionless function $f_{\rm KN}(b)$ is $f_{\rm KN}(b)=9g(b)/b^3$ with $b=4\gamma \varepsilon_{i}$, The factor $g(b)$ is expressed by
\begin{equation}
g(b)=(\frac{1}{2}b+6+\frac{6}{b})\rm ln(1+b)-(\frac{11}{12}b^3+6b^2+9b+4)\frac{1}{(1+b)^2}-2+2\rm Li_{2}(-b),\label{gfactor}
\end{equation}
where $\rm Li_2$ is the dilogarithm.

The emissivity of Compton scattering of directed photon beams on isotropically distributed electrons in solid angle $\delta \Omega_{\rm j}$ is given by
\begin{equation}
\delta L'_{ i,\nu'}={3c\sigma_{T} \over 16 \pi }\frac{h\nu'}{m_{\rm e} c^2}\int \int {u'_{ i} \over \varepsilon_{i}^2}{\delta N_{\rm \gamma}(z)\over \gamma^2}f(\gamma,\nu',\varepsilon_{i},\theta){\rm d}\varepsilon_{i} \rm d\gamma,
\end{equation}
where $\varepsilon_{i}=\varepsilon'/(2(1-\cos \theta)\gamma^2[1-\varepsilon'/\gamma])$, $\varepsilon'=h\nu'/m_{\rm e}c^2$, and
\begin{equation}
f(\gamma,\nu',\varepsilon_{i},\theta)=1+\frac{\varpi^2}{2(1-\varpi)}-\frac{2\varpi}{\tilde{b}(1-\varpi)}+\frac{2\varpi^2}{\tilde{b}^2(1-\varpi)^2},
\end{equation}
where $\tilde{b}=2(1-\cos \theta)\varepsilon_{i}\gamma$ and $\varpi=\varepsilon'/\gamma$.

The energy density of synchrotron emission field is given in Equation (\ref{SYND}). The energy density of the star and disk will be expressed as follows. The energy density of radiation field of the companion star is
\begin{equation}
u'_{\rm star} \simeq {\Gamma^{2} \over c}\int I_{\rm \nu',
star}(1-\beta_{\Gamma}\cos \theta_{\rm in})^{2}d\Omega,
\end{equation}
where $\theta_{\rm in}$ is the angle between the motion direction of the jet matter located at a three-dimensional element $\delta \Omega_{\rm j}$ and the trajectory of an incoming photon. $I_{\rm \nu'}$ is the intensity of the companion photon field that is expressed as
\begin{equation}
I_{\rm \nu'} = {L_{\rm star} \over 4 \pi R_{\rm star}^{2}}f_{\rm \nu'}(T_{\rm star}),
\end{equation}
where $L_{\rm star}$ and $R_{\rm star}$ are the luminosity and radius of the companion star, respectively. The normalized Planck function is given as
\begin{equation}
f_{\rm \nu'}(T_{\rm star}) = {2\pi h \nu'^{3}\over \sigma_{\rm SB} c^{2}T_{\rm star}^{4}}{1\over \exp({h\nu' \over k
T_{\rm star}})-1},\label{fstar}
\end{equation}
where $T_{\rm star}$ is the temperature of the companion star, and $\sigma_{\rm SB}$ is the Stefan-Boltzmann constant.

The spectra of an accretion disk in the low-hard states of black hole X-ray binaries is quite involved (see also \citealt{ZhangX13}). It can be assumed that a standard thin accretion disk is truncated at a certain radius to an optically thin, geometrically thick accretion disk.  To obtain properties of radiation field of the accretion disk, we use approximated expressions from Equations (1) - (4) of \cite{ZAA09}, which is updated by using renewed fitting parameters (\citealt{ZAA14b}). The energy densities of radiation field of the accretion disk is
\begin{equation}
u'_{\rm disk}=\frac{9d^2EF_{\rm in}(E)}{4R_{\rm in}^2c}+\frac{9d^2EF_{\rm out}(E)}{4R_{\rm out}^2c},
\end{equation}
where $d$ is the orbital radius of the system. $R_{\rm out} = 10^{11}\ \rm cm$  is the outermost truncation radius of the disk and $R_{\rm h} = 10^{8}$ cm is the radius of the inner X-ray emitting source. The intensity of photons of the disk is given by
\begin{equation}
F_{\rm disk}(E) = F_{\rm in}(E)+F_{\rm out}(E)= { K_{\rm X}({E_{\rm b} \over  1\ \rm keV})^{-\alpha}
\exp({- {E \over E_{\rm c}}}) \over ({E \over E_{\rm b}})^{-2}+({E
\over E_{\rm b}})^\alpha } + {K_{\rm d}({E_{\rm out} \over
1\ \rm keV})^{1 \over 3}\exp({-{E \over E_{\rm b}}})\over ({E \over E_{\rm
out}})^{-2} + ({E \over E_{\rm out}})^{-{1\over 3}}},
\end{equation}
where $K_{\rm X} = 1.1$ $\rm cm^{-2}$ $\rm s^{-1}$, $E_{\rm b} = 0.4$ $\rm keV$, $E_{\rm c} = 140$ $\rm keV$, $\alpha = 0.44$, $E_{\rm
out} = ({R_{\rm out} \over R_{\rm h}})^{-{3 \over 4}}E_{\rm b}$. $K_{\rm d}$ can be obtained from the condition that the inner and outer components
intersect at $E_{\rm i} = 1.1$ $\rm keV$. Concerning relevant details, interested readers are referred to \cite{ZAA09} and \cite{ZAA14a}

\subsection{Observed Spectra}

The monochromatic luminosity produced at a given height that is a superposition of radiation emitted at each three-dimensional element, and is given as
\begin{equation}
L^{\rm II}_{\nu _{\rm obs}}(z) = \sum {\partial \delta L_{\nu_{\rm
obs}}(z,\theta) \over \partial \Omega_{\rm obs}}=\mathfrak{D}^3\sum {\partial \delta L_{\nu'}(z,\theta') \over \partial \Omega'_{\rm obs}}, \label{unLum}
\end{equation}
where
\begin{equation}
 {\nu_{\rm obs}= \nu^{'} \mathfrak{D}};\ \ \ \ \ z =z_{\rm 0}+ {(z-z_{\rm 0}) \over 1-\beta_{\Gamma}\cos \theta};\ \ \ \ \ \cos \theta = {\cos \theta^{'}+\beta_{\rm \Gamma} \over 1+\beta_{\rm \Gamma} \cos \theta^{'}};
\end{equation}
and the Doppler factor depends on the bulk Lorentz factor of the jet $\Gamma$ and the viewing angle $\theta$. As for an approaching jet, we have
\begin{equation}
\mathfrak{D} = \Gamma(1+\beta_{\rm \Gamma}\cos \theta^{'}) = {1 \over
\Gamma (1-\beta_{\rm \Gamma}\cos \theta)}. \label{Doppler}
\end{equation}

In the case of the X-ray binary Cygnus X--1, the dominated radiation field of the stellar companion provides a suitable target for an absorption of high energy photons. Therefore, we also involve electromagnetic cascade processes due to $\gamma$-$\gamma$ interactions. Regarding the calculation of optical depth, some literature on X-ray binaries can be found (\citealt{Bottcher05, Dubus06,Cerutti11}). Following these investigations, we consider an anisotropic absorption process of the photon fields of the companion star. Because of emission region away from the central black hole, we find that there are the same results via taking the finite size (\citealt{Dubus06}) or the point-like source approximation (\citealt{Bottcher05}) of the stellar companion into account. After considering absorption processes of $\gamma$-$\gamma$ interactions, we have the attenuation of the production luminosity
\begin{equation}
L_{\nu _{\rm obs}}^{\rm I}(z) = L^{\rm II}_{\nu _{\rm obs}}{\rm exp}[-\tau(z)],
\end{equation}
where $\tau(z)$ is the optical depth of $\gamma$-$\gamma$ interactions. The spectrum of pairs produced in $\gamma$-$\gamma$ interaction processes is given as
\begin{equation}
L_{\nu,{\rm pairs}}^{\rm I}(z) = L^{\rm II}_{\nu _{\rm obs}}\{1-{\rm exp}[-\tau(z)]\} .
\end{equation}
These pairs will excite electromagnetic cascade to produce high energy photons. The total output luminosity from the whole emission region located in a jet can be obtained by summing $L^{\rm I}_{\nu _{\rm obs}}(z)$ from $z_{\rm 0}$ to $z$, that is
\begin{equation}
L_{\nu _{\rm obs}} = \sum L^{\rm I}_{\nu _{\rm obs}}(z), \label{LZSUM}
\end{equation}
where $z$ is the height of the jet from the central black hole, $z_{\rm 0}$ corresponds to the onset of the emission region.

\subsection{Numerical Methods}
Equation (\ref{dNdz}) of electron evolution is solved with the implicit difference scheme based on the methods used in \cite{CG99} and \cite{ms03}. The uniformly spaced logarithmic energy grid, with intervals $\bigtriangleup\gamma_{i} = \gamma_{i+{1\over 2}} -\gamma_{i-{1\over 2}}$, is set as
\begin{equation}
\gamma_{i} = \gamma_{\rm min}({\gamma_{\rm max}\over \gamma_{\rm min}})^{i-1\over
m-1}\ ; \ \ \ \ \ \  (i = 0, ...,  m).
\end{equation}
Adopting a definition of $N^{j}_{i} =N_{\rm \gamma}(\gamma_{i}, j\Delta r)$, the evolution Equation (\ref{dNdz}) can be differenced as linear equations
\begin{equation}
U^{j}_{i}=A_{i}N^{j+1}_{i-1} + B_{i}N^{j+1}_{i} + C_{i}N^{j+1}_{i+1},
\label{ABCU}
\end{equation}
where the coefficients are
\begin{equation}
A_{i} = 0;\ \ \ \ \ \   B_{i} = 1 + ({d\gamma\over dz})_{i-{1\over 2}}{\Delta z\over \Delta
\gamma_{i}} + {\Delta z\over z_{\rm max} - z_{\rm 0}}; \ \ \ \ \ \ C_{i} = -({d\gamma\over dz })_{i+{1\over 2}}{\Delta z\over \Delta
\gamma_{i}}.
\end{equation}
The non-homogeneous term is expressed as
\begin{equation}
U^{j}_{i} = N^{j}_{i} + {\Delta z\over
c\beta_{\Gamma}\Gamma}Q^{j}_{i}.
\end{equation}

We solve numerically Equation (\ref{ABCU}) by employing a tridiagonal routine code (\citealt{Press07}). At different heights of the jet, this equation is solved iteratively. Initially, the energy density of synchrotron emission is set to be $u_{\rm Syn}'=0$ and the electron distribution is calculated using this initial value. Subsequently, this electron distribution is used to calculate synchrotron emission density. The new value of synchrotron emission density is used to recalculate the electron distribution. This process is repeated until convergence. Furthermore, we use the Gauss integral method to carry out numerical calculations, which can efficiently speed up calculation processes.

\section{Numerical Results}
Cygnus X--1 is a confirmed high mass microquasar at a distance of $D = 1.86\ \rm kpc$ (\citealt{Reid11}), with an orbital period of $P\simeq 5.6$ days. We use the following updated parameters of this system following \cite{Ziolkowski14}. The mass of central black-hole is $M_{\rm BH}\simeq16\ \rm M_{\odot}$, where $\rm M_{\odot}$ is the solar mass. The radius of the companion star is $R_{\rm star }\simeq19$ $\rm R_{\odot}$, where $\rm R_{\odot}$ is the solar radius, and its effective temperature is $T_{\rm star}\simeq2.8\times10^{4}\ \rm K$. The average orbital radius of this binary is $d\simeq3.2\times 10^{12}\ \rm cm$, and the inclination of the normal to the orbital plane with respect to the line of sight is $\varphi_{\rm obs}=29^{\circ}$. Furthermore, we adopt the bulk Lorentz factor of the jet of $\Gamma=1.25$ (e.g. \citealt{ZAA14b}), and the half-opening angle of the jet of $\varphi_{\rm j} = 1^{\circ}$ (\citealt{Stirling01,ZAA14b}).

\begin{figure*}[]
  \begin{center}
  \begin{tabular}{ccc}
\hspace{-0.79cm}
     \includegraphics[width=80mm,height=80mm]{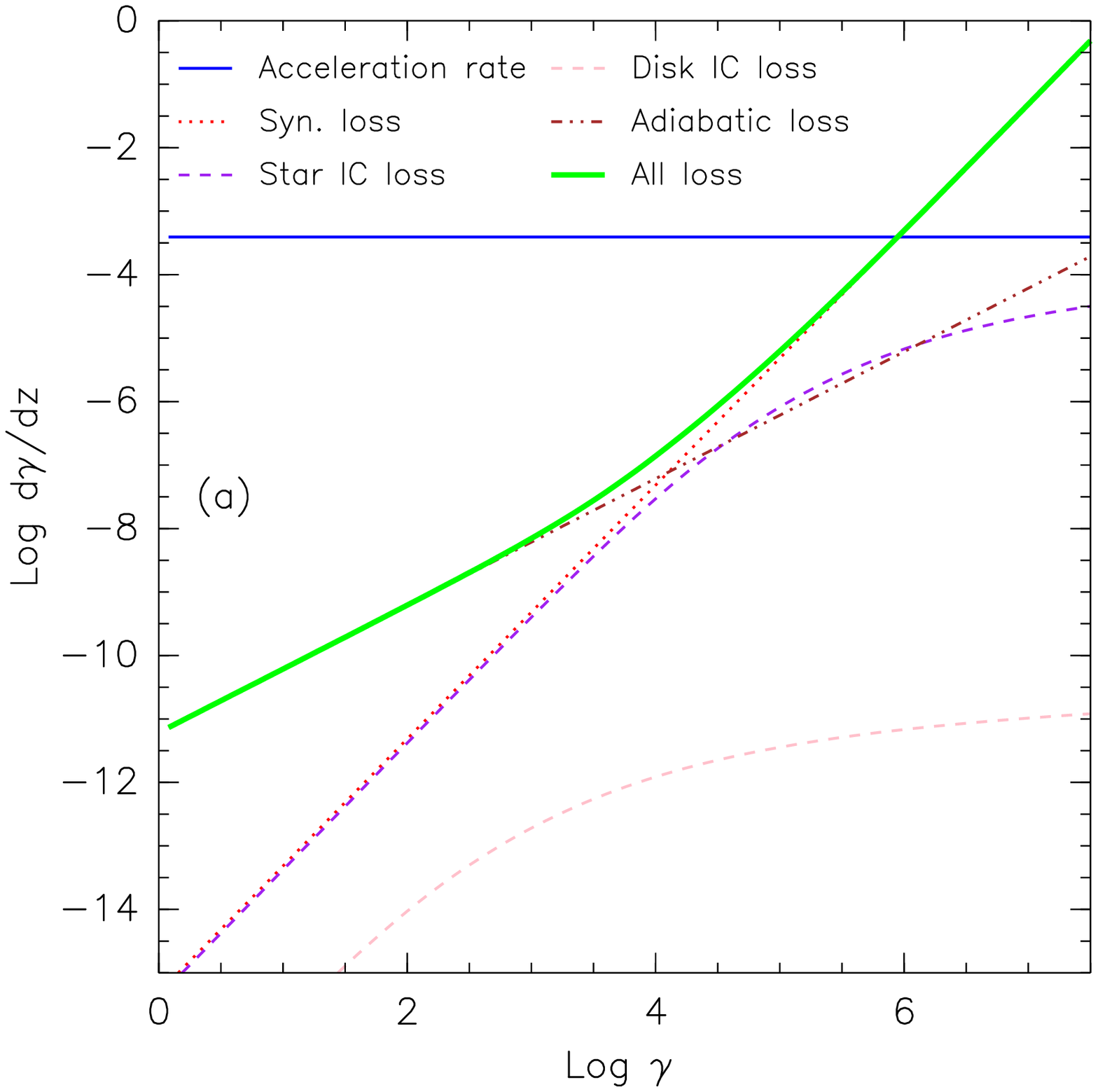} & \ \ \
\hspace{-0.79cm}
     \includegraphics[width=80mm,height=80mm]{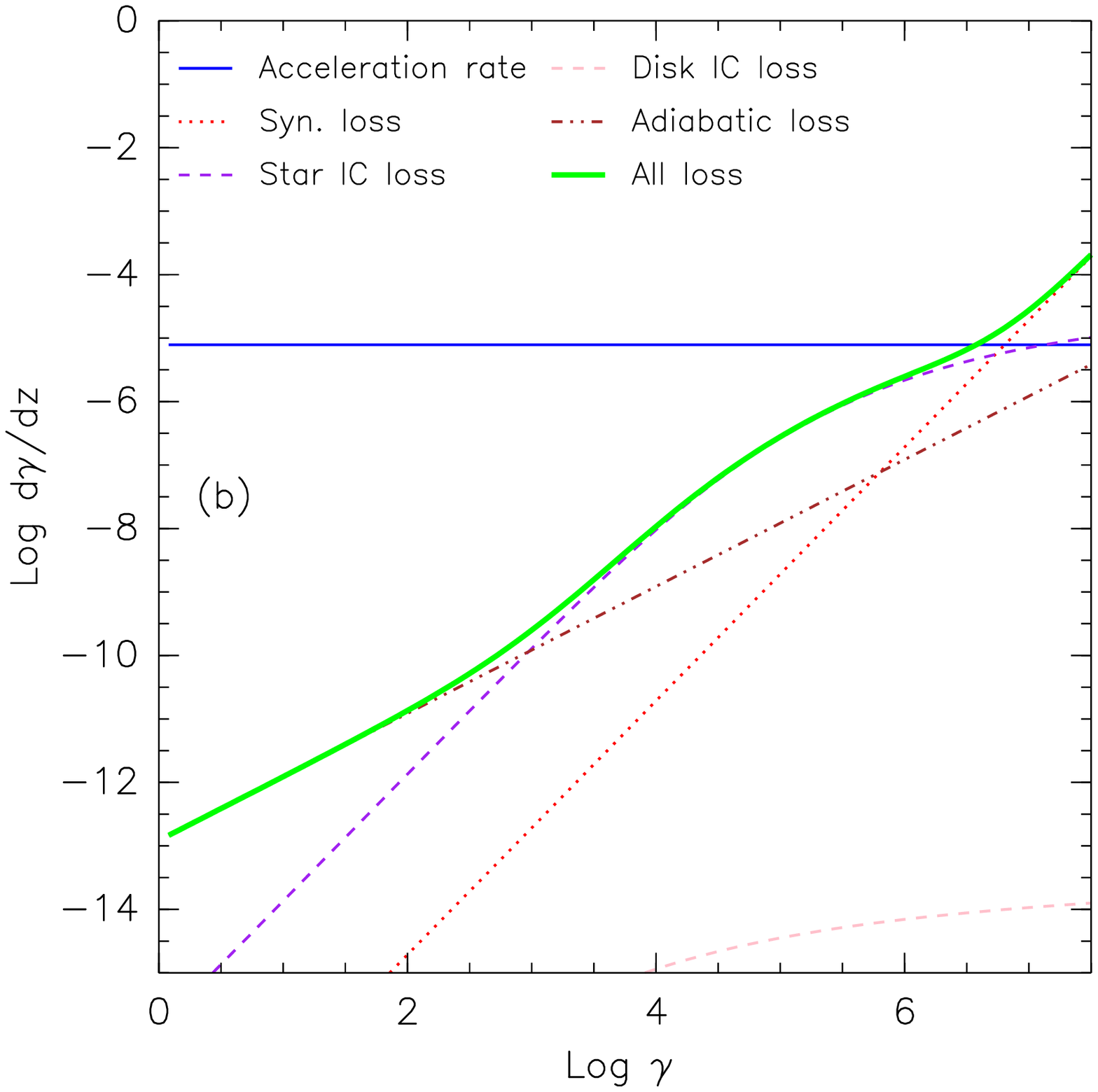}
\end{tabular}
  \end{center}
\caption{Acceleration and cooling rates at the height of the jet $z=1.0\times 10^{11}\ \rm cm$ (panel a) and $5.0\times 10^{12}\ \rm cm$ (panel b), calculated for the model parameters $\eta=0.005$ and $B_{\rm 0}'=1.0\times10^2\ \rm G$. Point of intersection of the thick and thin solid lines corresponds to the maximum energy of relativistic electrons.} \label{figs:losses}
\end{figure*}

\begin{figure*}[]
  \begin{center}
  \begin{tabular}{ccc}
\hspace{-0.79cm}
     \includegraphics[width=80mm,height=70mm]{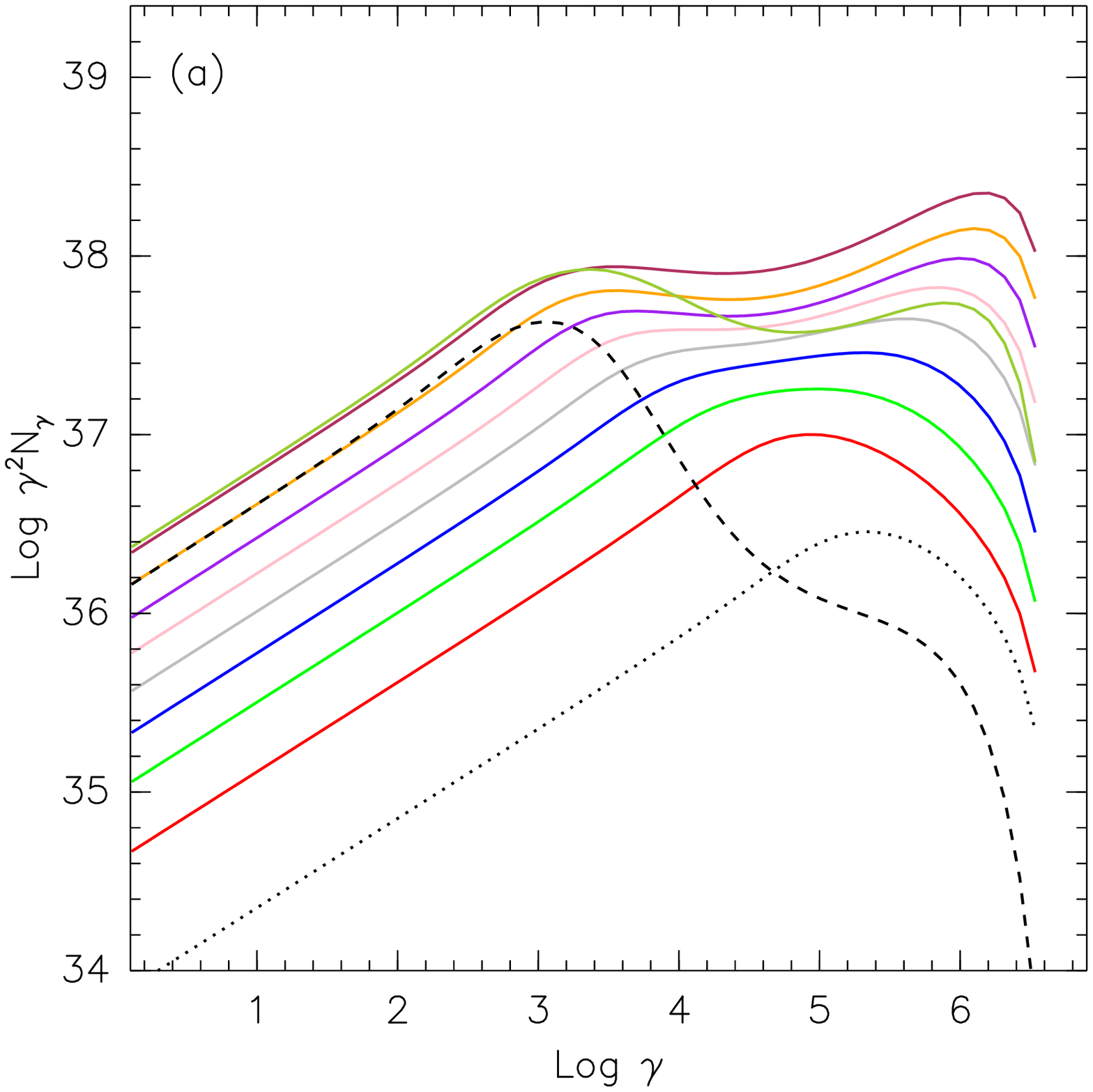}& \ \ \
\hspace{-0.79cm}
     \includegraphics[width=80mm,height=70mm]{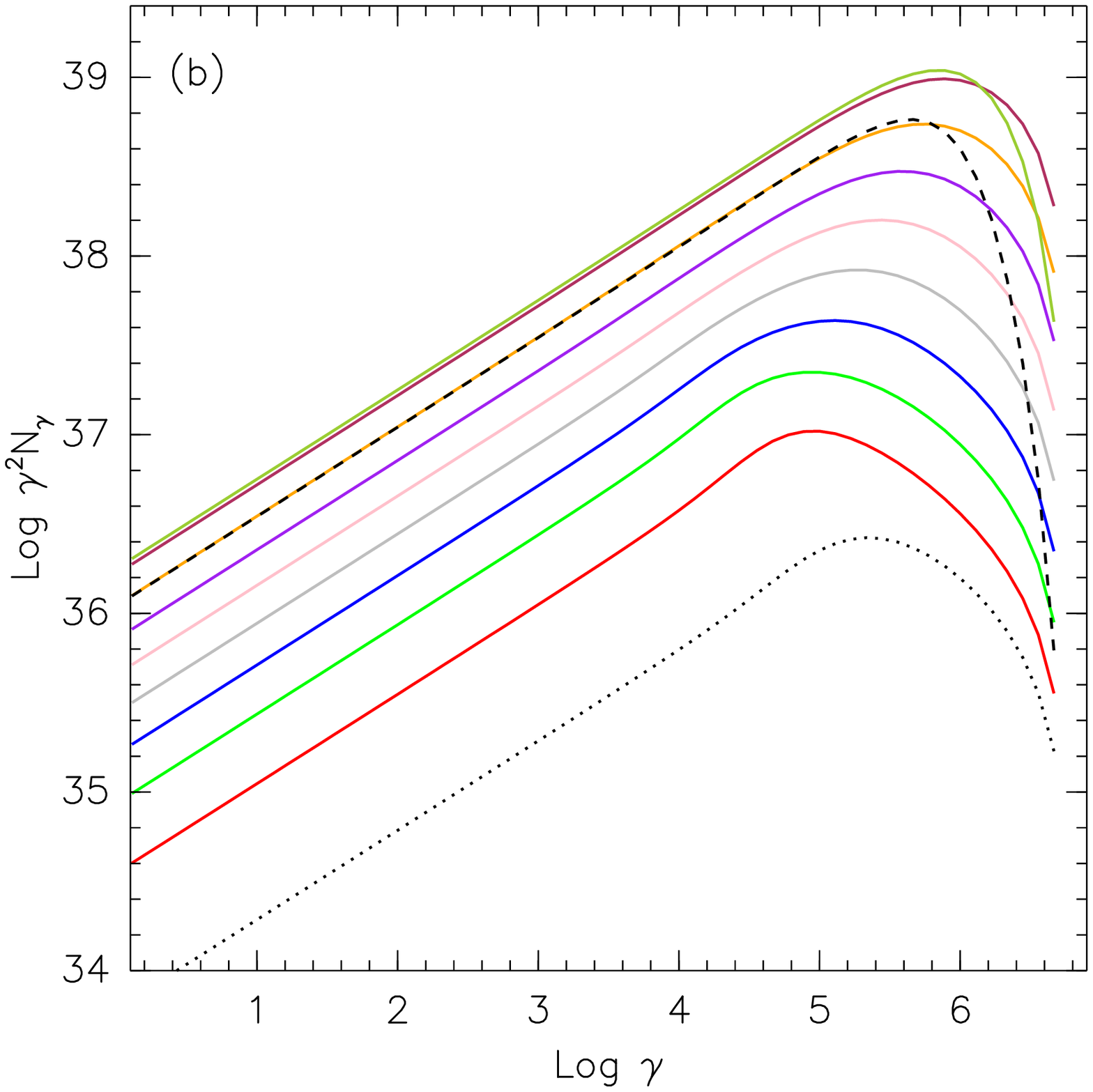}\\
\hspace{-0.79cm}
     \includegraphics[width=80mm,height=70mm]{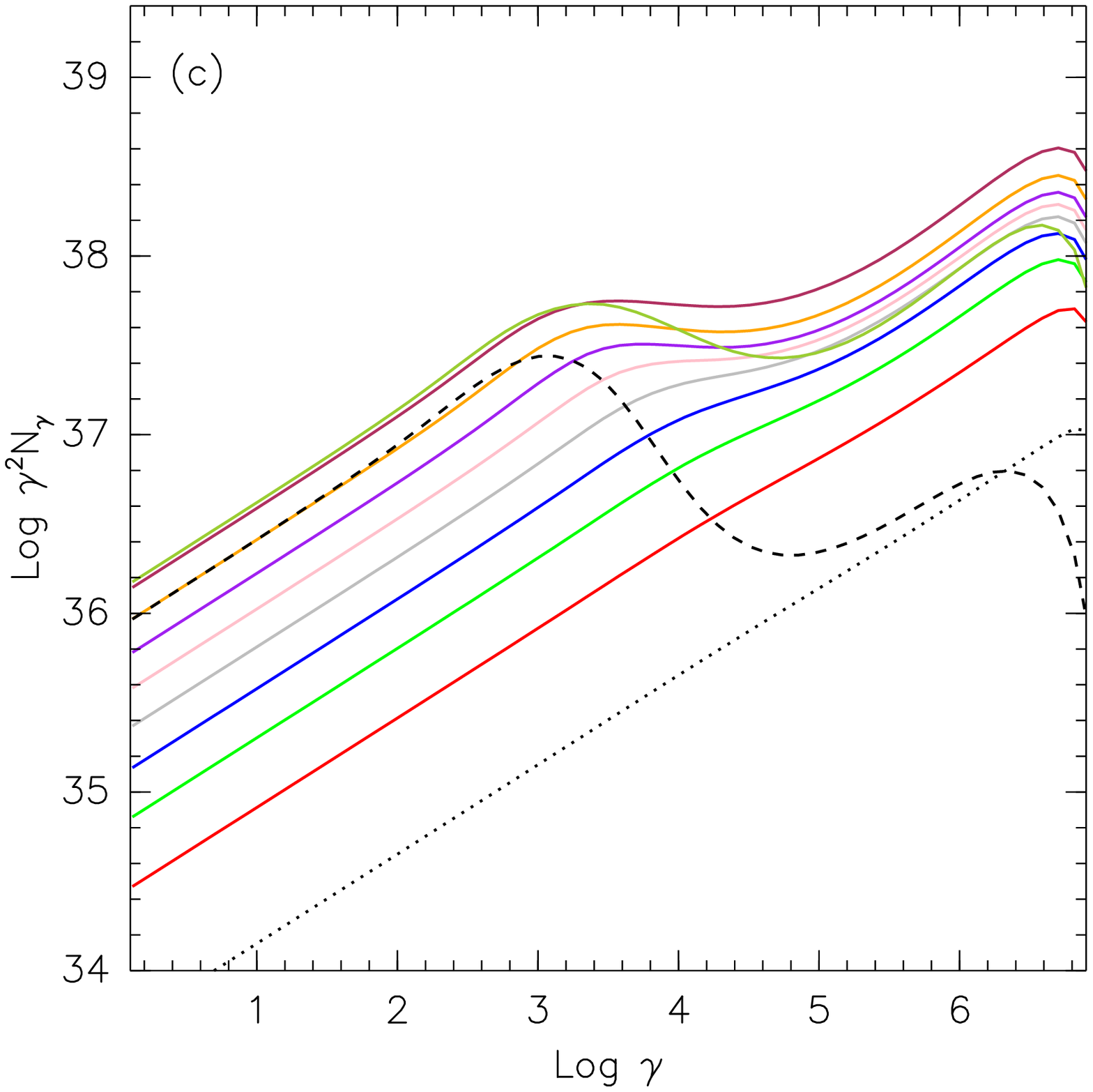}&\ \ \
\hspace{-0.79cm}
     \includegraphics[width=80mm,height=70mm]{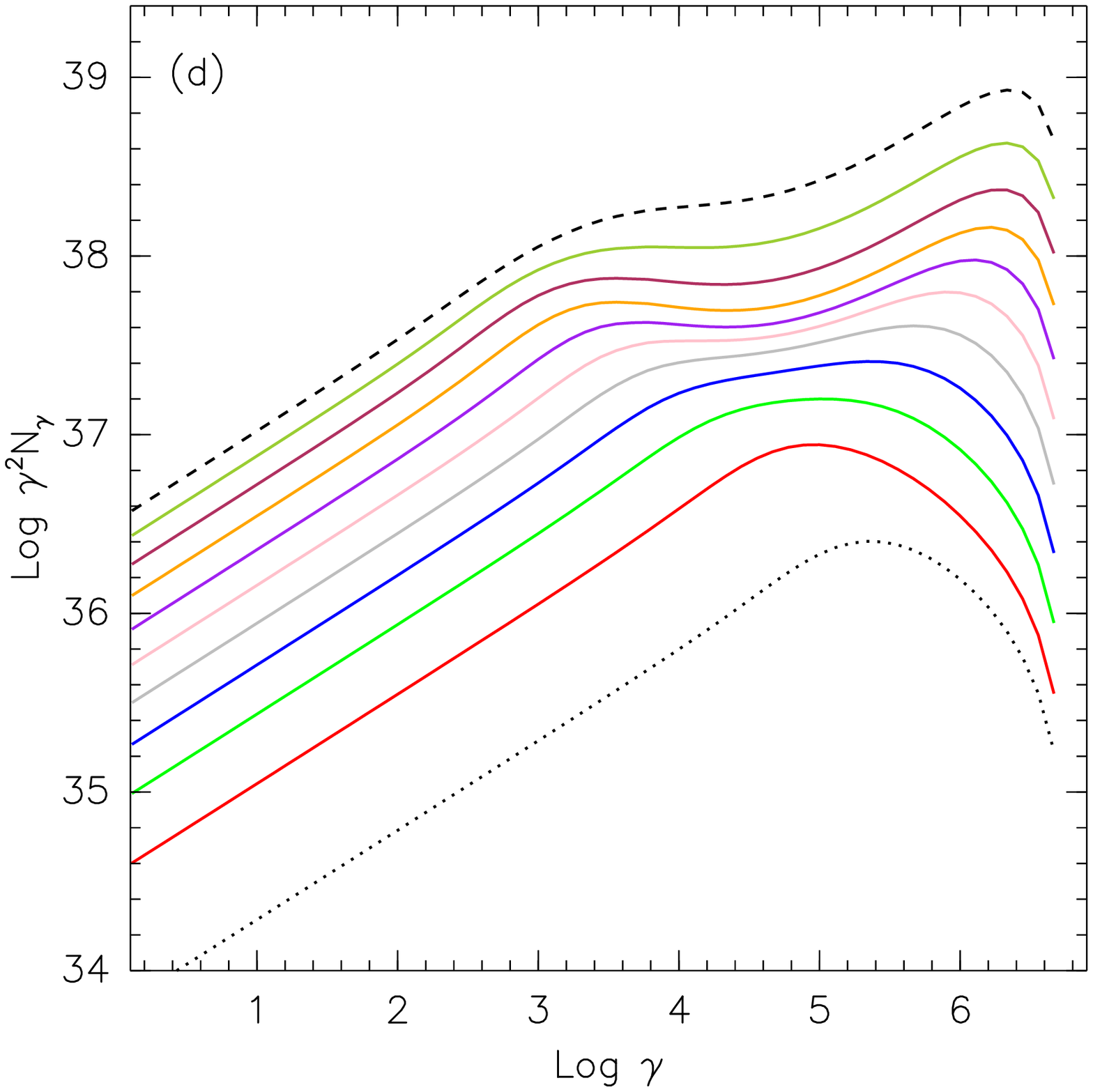}& \ \ \
\end{tabular}
  \end{center}
\caption{Time evolution of spectra of relativistic electrons. The explored evolution zone is the case that the electron injection starts at $z_{\rm 0}=1.0\times10^{11}\ \rm cm$, and ends at $z_{\rm inmax} =5.0\times 10^{12}\ \rm cm$ (for panels a--c) and $1.0\times 10^{13}\ \rm cm$ (for panel d). Panels (a) and (d) include all radiative losses mentioned in this work. Panel (b) involves synchrotron and its Comptonization and panel (c) only involves external inverse Compton processes. The adopted parameters are $B_{\rm 0}^{'} = 1.0\times10^{2}\ \rm G$, $\gamma_{\rm min} = 1$,  $\eta=0.005$,  $p=1.5$ and $q_{\rm rel} = 0.01$. Electron spectral distributions are plotted for various explored scenarios in a logarithmic step of 0.2 between the curves. The dotted, dashed lines on each panel indicate the start and end of electron injection, respectively.}  \label{figs:ElectronEvolution}
\end{figure*}

\subsection{Electron Maximum Energy}

The maximum energy of electron is determined by the balance of the cooling rates (including synchrotron emission loss, inverse Compton scattering losses for star and disk, and adiabatic loss) and acceleration rate. For a standard first order \emph{Fermi} acceleration process, the acceleration rate for a particle of energy $\gamma$ in a magnetic field $B(z)$ is given by
\begin{equation}
{{\rm d}\gamma \over {\rm d}z}=\frac{\eta eB(z)c}{m_{\rm e}c^2}\frac{1}{c\beta_{\rm \Gamma}\Gamma}\ , \label{gain}
\end{equation}
where $e$ is the charged charge of an electron and $\eta$ is a parameter that characterizes the acceleration efficiency in a dissipation region. Generally, one should solve the non-linear equation of particle energy equilibrium in order to obtain the maximum energy of electrons $\gamma_{\rm max}(z)$ that is a function of the height of the jet $z$.

Through considering the balance of the acceleration rate and all cooling rates, we can obtain the maximum energy of accelerated electrons. As a demonstration, we show in Figure \ref{figs:losses} acceleration and cooling rates at the height of the jet $z=1\times 10^{11}\ \rm cm$ (panel a) and $5\times 10^{12}\ \rm cm$ (panel b). The model free parameters $\eta=0.005$ and $B_{\rm 0}'=1\times10^2\ \rm G$ are used in this figure. The maximum energies $\gamma_{\rm max}=8.32\times10^{5}$ (for panel a) and $2.51\times10^{6}$ (for panel b) of relativistic electrons can be derived from the intersection of the thick and thin solid lines. Close to the base of the jet, the existence of a strong magnetic field is responsible for an efficient synchrotron emission loss. At larger distance from the black hole (close to the binary system scale), the synchrotron emission loss decreases due to decreasing magnetic field intensity, where the inverse Compton scattering from the stellar companion dominates energy loss of electrons. It is should note that, at the binary system scales, the synchrotron emission loss is still important provided that a more higher magnetic field strength is set at the base of the jet. When reaching very large distance (close to the termination of the jet), the adiabatic loss is a dominant cooling mechanism. Adopting the same method, we can obtain the maximum energy of accelerated electrons. In the case, the maximum energy is very high and over expected limits, which has no physical meaning. Therefore, we have to assume that the acceleration process is impeded at a certain height of the jet, in which a possible maximum energy ($\gamma_{\rm max}\sim10^{7}$) can be obtained and the injection of primary electrons is also stopped.

\subsection{Electron Spectral Evolution}

To investigate evolution properties of relativistic electrons, we fix a zone between $z_{\rm 0}=1.0\times10^{11}\ \rm cm$  and $z_{\rm max}=1.0\times10^{13}\ \rm cm$ in the jet, in which evolution spectra of the relativistic electrons for Cygnus X--1 are presented in Figure \ref{figs:ElectronEvolution}. Throughout the paper, two parameters $q_{\rm jet}=0.1$ and $\dot{M}_{\rm acc}=10^{-9} M_{\rm \odot}\  \rm yr^{-1}$ are fixed. The magnetic field strength at the height $z_{\rm 0}$ in the jet frame is assumed to be $B_{\rm 0}^{'} = 1\times10^{2}\ \rm G$, and the acceleration efficiency to be $\eta=0.005$. The electron-related parameters are: the minimum energy of $\gamma_{\rm min} = 1$ and its spectral index of $p=1.5$, the conversion efficiency of the jet matter of $q_{\rm rel} = 0.01$. Electron spectra are presented for various scenarios explored in a logarithmic step of 0.2 between the curves.

In panel (a) of Figure \ref{figs:ElectronEvolution}, we consider a case of the time evolution of relativistic electrons, in which the injection of electrons is stopped at the height $z_{\rm inmax}=5\times 10^{12}\ \rm cm$, corresponding to the maximum Lorentz factor of $2.51\times10^{6}$. The various radiative cooling mechanisms involved in this work are included in this panel. At the beginning of the injection of electrons, the number of relativistic electrons increases with the height of the jet. In the later stages of evolution, i.e, reaching the distance of the orbital radius, electron spectra present some shallow troughs as a result of competition between synchrotron emission cooling and dominant inverse Compton scattering loss in the Klein-Nishina regime from the stellar companion. After stopping the injection of electron, electron spectra at high energy bands decay rapidly, whereas the number of the low energy electron ($\gamma < 10^3$) reaches saturation and then decreases slightly. The main reason is that the low energy electron population injected have a relativistically inefficiently radiative cooling process, whereas high energy electrons injected can efficiently cool by the radiative processes mentioned. Below, we are to explore the time evolution behavior of electron spectra for some important cooling components.

The influence of synchrotron and its self Compton scattering cooling regarding electron spectra is studied in panel (b). We see that the number of high energy electrons increases continuously with the peak frequency at $\gamma \sim 10^6$, more rapidly than other situations (see panels a, c and d), until the height where electron injection is stopped. After the injection of electrons is stopped, electron spectra decrease with increasing height as a result of without any injection of fresh electrons. In particular, these spectra decay very rapidly at high energy tail ($\gamma >10^6$) due to the dominant synchrotron loss, while spectra at low energy bands reach saturation and then decrease slowly due to inefficient cooling process for low energy electrons. In panel (c), we present the case of considering external inverse Compton scatterings, i.e., stellar companion and accretion disk. As shown in this panel, the number of relativistic electrons, with a high energy cut-off spectrum at $\gamma >10^6$, increases with increasing height, appearing some troughs at high energy bands as a result of the radiative losses in the Klein-Nishina regime. Similarly, the spectra at the high energy bands decrease fast when electrons stop injecting. We study the influence of processes of electron injection on its spectra in panel (d). In this case, relativistic electrons are injected consistently through the considered whole zone along the jet. The number of relativistic electrons constantly increases with increasing height, as shown in this panel.

Except for the effects of radiative coolings regarding electron spectra, we will discuss briefly the influence of adiabatic and escape losses and electron spectral index on its spectra. The adiabatic loss is important for the cooling of low energy electrons (see also Figure \ref{figs:losses}). Furthermore, these low energy electrons can also escape the emission region due to having a much longer radiative cooling timescale. When taking a soft electron spectral index into account, relativistic electrons have a strong cooling process. Spectra at high energy bands present a power-law plus an exponential high energy cut off. Meanwhile, the cooling of the low energy electron population is also more efficient.

\begin{figure*}[]
  \begin{center}
  \begin{tabular}{ccc}
\hspace{-0.79cm}
     \includegraphics[width=80mm,height=80mm]{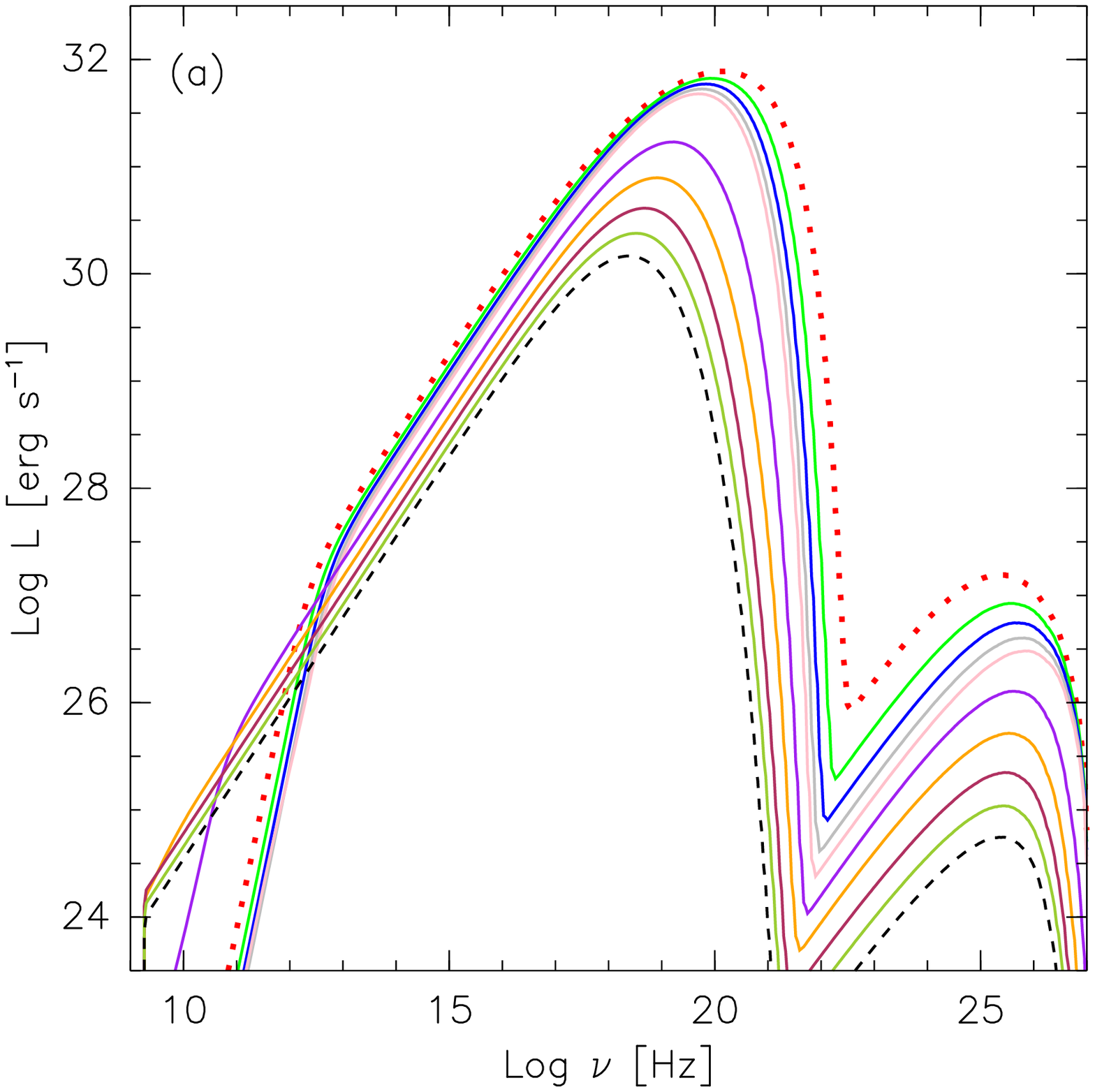} & \ \ \
\hspace{-0.79cm}
     \includegraphics[width=80mm,height=80mm]{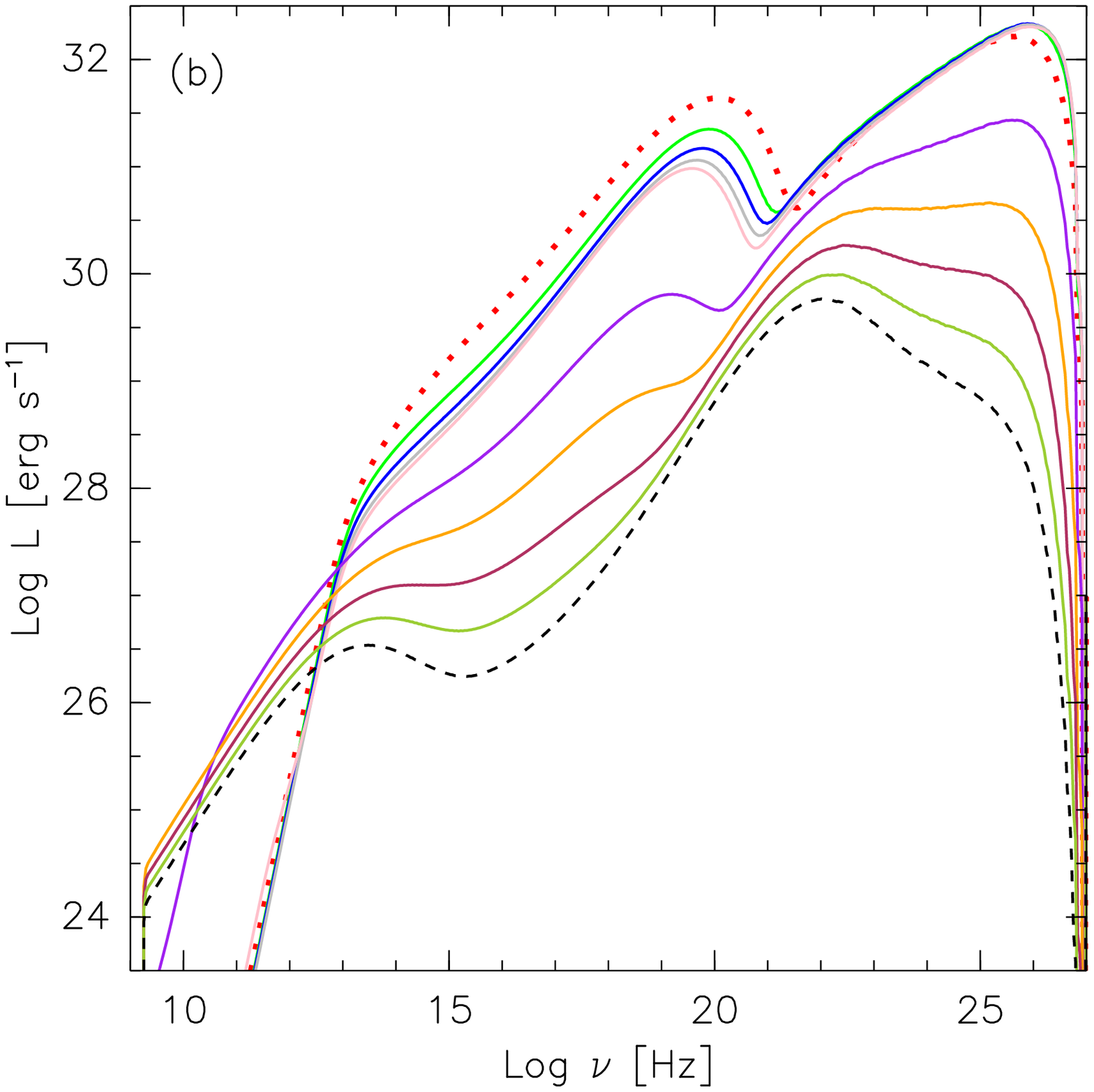}
\end{tabular}
  \end{center}
\caption{Spectral energy distributions of photons at different heights of the jet. A logarithmic step of 0.2 is used between the plotted curves. The adopted parameters are $z_{\rm 0}=1.0\times10^{11}\ \rm cm$, $z_{\rm inmax} =5.0\times 10^{12}\ \rm cm$, $z_{\rm max}=1.0\times10^{13}\ \rm cm$, $B_{\rm 0}^{'} = 3\times10^{2}\ \rm G$, $\gamma_{\rm min} = 1$, $\eta=0.005$, $p=1.5$ and $q_{\rm rel} = 0.01$.  Panel (a) incudes synchrotron emission and its Comptonization. Synchrotron emission, SSC and external inverse Compton scattering processes of the stellar companion and accretion disk are involved in panel (b). The dotted, dashed lines on two panels indicate start and termination locations of the emission region.} \label{figs:spectra}
\end{figure*}

\begin{figure*}[]
  \begin{center}
  \begin{tabular}{ccc}
\hspace{-0.79cm}
     \includegraphics[width=60mm,height=70mm]{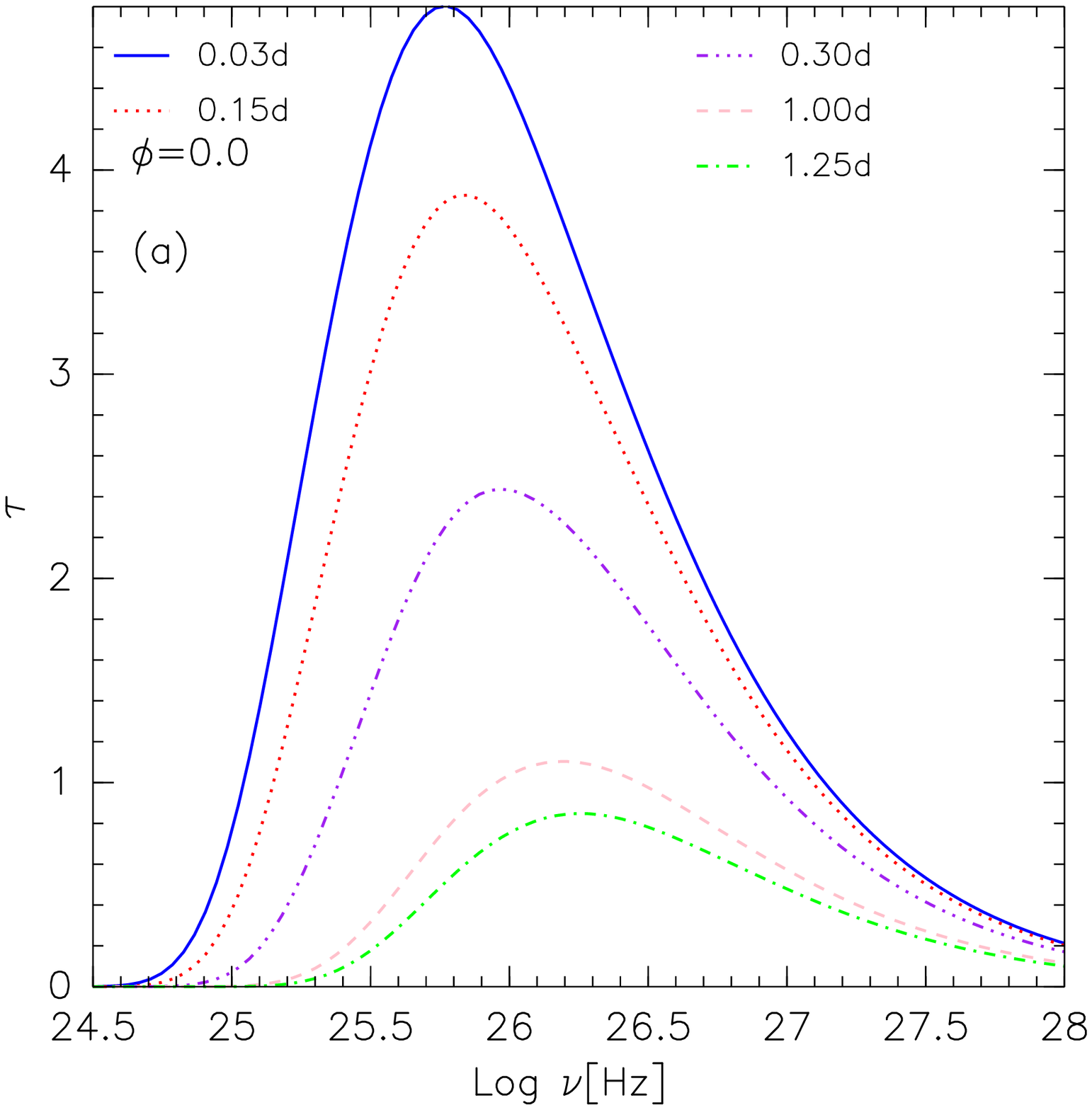} & \ \ \
\hspace{-0.79cm}
     \includegraphics[width=60mm,height=70mm]{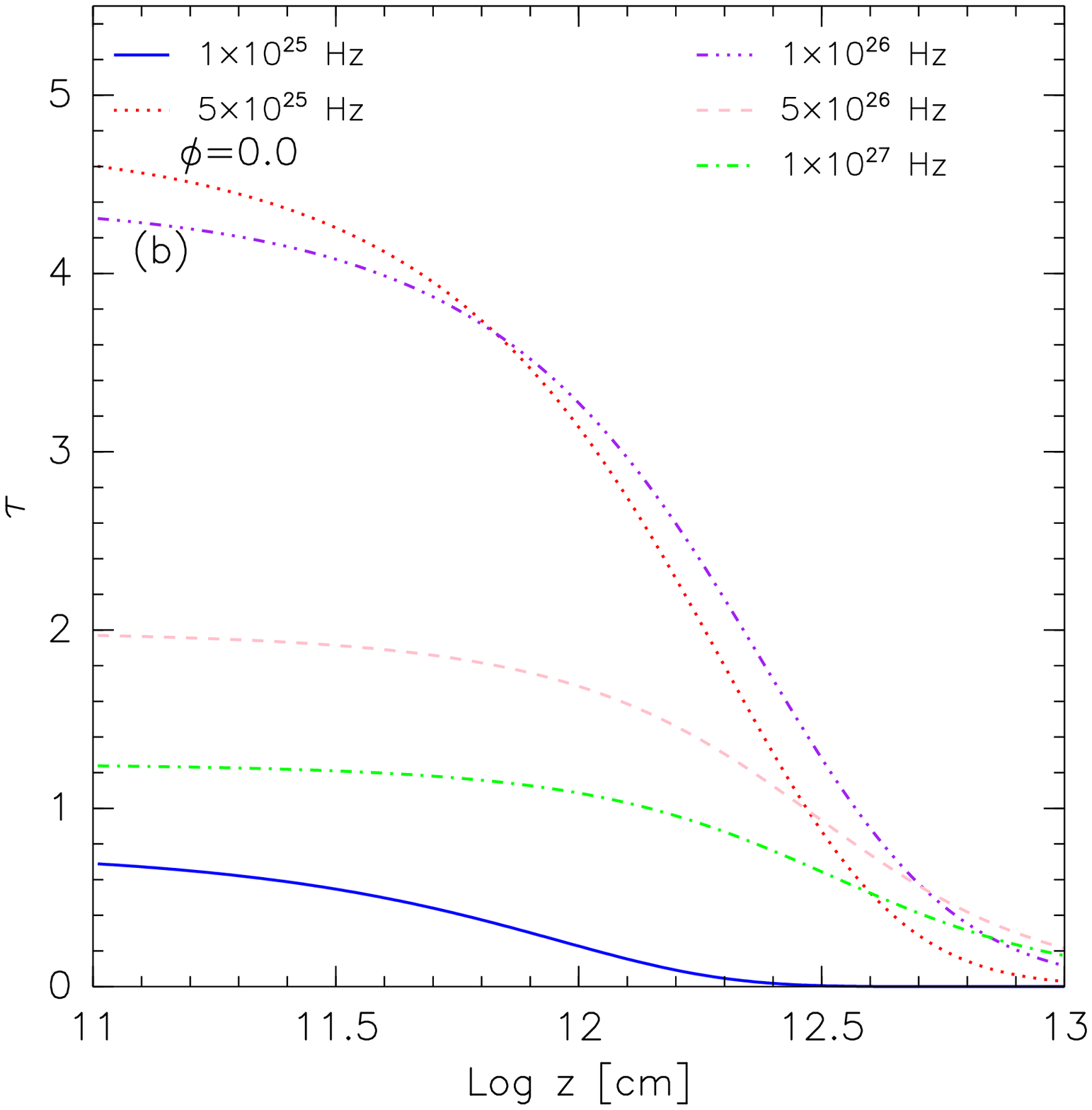}& \ \ \
\hspace{-0.79cm}
     \includegraphics[width=60mm,height=70mm]{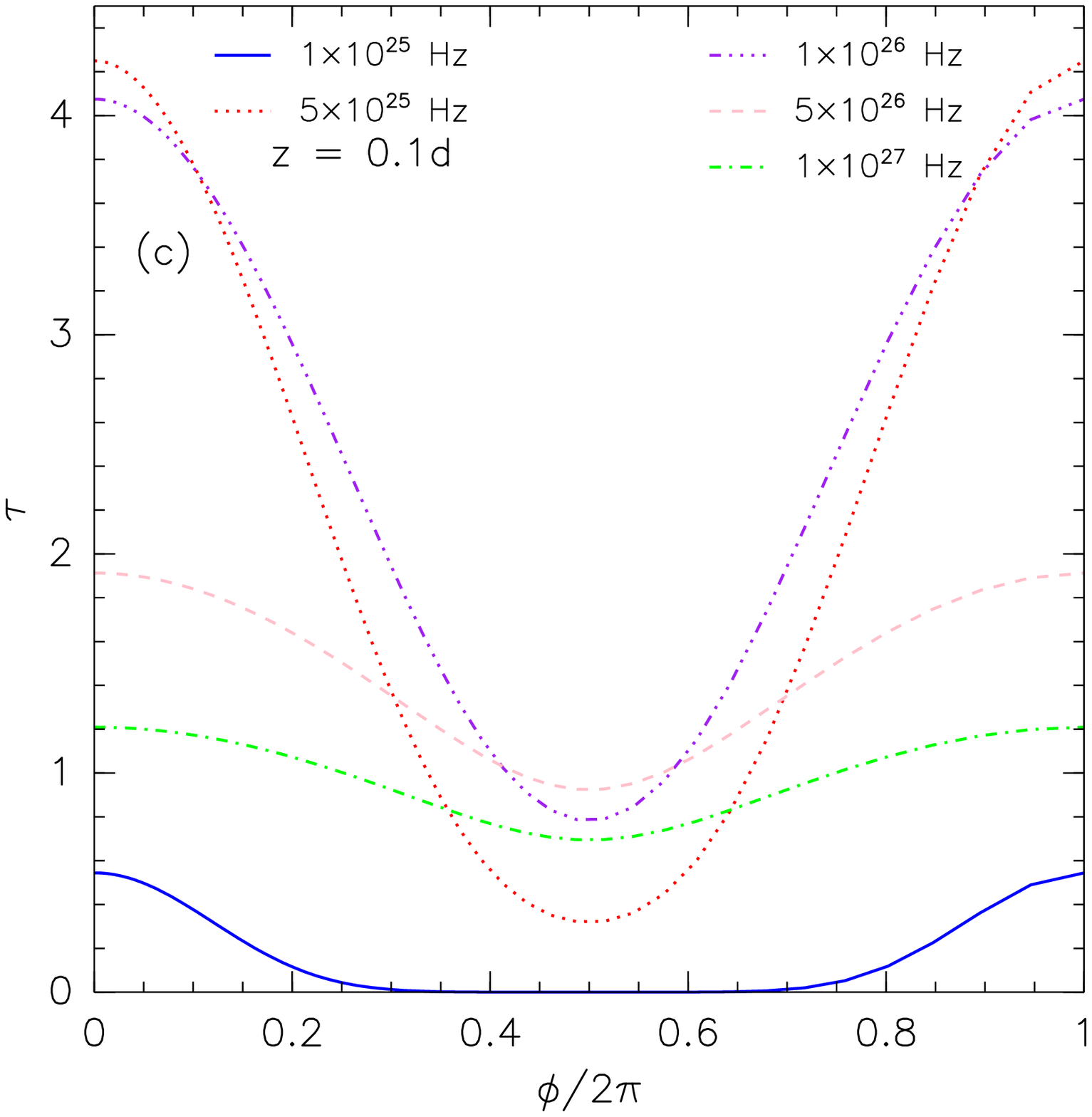}
\end{tabular}
  \end{center}
\caption{The optical depth of $\gamma$-$\gamma$ interactions as the functions of photon energy (panel a), height of the jet (panel b) and orbital phase of the system. In panel (a), $d$ is the orbital radius of the system.}  \label{figs:tau}
\end{figure*}

\subsection{Photon Spectra}

We turn now to investigate the properties of photon spectra from Cygnus X--1. The sum of each individual spectrum at different heights of the jet are responsible for the average emission spectra observed (see also Equation \ref{LZSUM}). Each spectrum produced at a different distance of the emission region are presented in Figure \ref{figs:spectra}. In order to highlight main aspects, we do not include an absorption effect of $\gamma$-$\gamma$ interactions and cascade processes in this figure. The used parameters are similar to those of panel (a) of Figure \ref{figs:ElectronEvolution}. Step between the plotted spectra is a logarithmic step of 0.2. The dotted, dashed lines indicates spectra at the start and end locations of the emission region.

As shown in Figure \ref{figs:spectra}, synchrotron self absorption is very evident at the base region of the jet and then this effect gradually weakens with increasing  height of the jet. Spectral energy distributions shown in panel (a) involve synchrotron emission and its Comptonization. The first of two peak frequencies at low energy bands, having strong emission fluxes, is from the synchrotron emission, and the other peak from the SSC process. With increasing heights of the jet, these peaks shift toward lower energy bands and emission fluxes also reduce. In particular, the shift of the first peak is very obvious due to the decreasing of magnetic field strength. The spectra in panel (b) include various radiative processes such as synchrotron emission, SSC and external inverse Compton scattering processes from the stellar companion and accretion disk. At the beginning of electron injections, the spectra appear also two peak frequencies at about $10^{20}\ \rm Hz$ and $10^{26}\ \rm Hz$, respectively. The first is due to synchrotron emission of relativistic electrons, and the second external inverse Compton scattering of these electrons off photons of the star (dominated) and accretion disk. From this panel, we see that the second peak (at about $10^{26}\ \rm Hz$) firstly almost changes, and then decreases owing to decreasing the photon density of the companion. Furthermore, the first peak decreases constantly and then disappears as a result of the superposition of a weak synchrotron emission and a strong inverse Compton scattering. Finally, a third low flux peak at about $10^{13}\ \rm Hz$ appears; this is because both emission intensities are inverted. These individual spectra at different heights will contribute to the total photon spectra by summing them.

By analyzing individual photon spectral distributions at different heights of the jet, we can know that radio emissions come mainly from the outer part of the emission region. Nevertheless, the peaks of synchrotron emission and inverse Compton scattering are from the base region of the emission region. The exponential cutoff part (over the peak frequency) of the total synchrotron spectra originates from the innermost part of the emission region. According to the fitting to observations, one can infer the location which different band observations are produced at.

\subsection{Absorption Process}
In the high mass microquasar Cygnus X--1, the photon field of its stellar companion has an important impacts regarding radiation properties of the system.  The optical depth of $\gamma$-$\gamma$ interactions as the functions of energy, height and phase angle are presented in Figure \ref{figs:tau}. The orbital phases are defined to be $\phi=0$ at the superior conjunction, and $\pi$ at the inferior conjunction.

At the superior conjunction, panel (a) shows optical depth as a function of energy for different heights of the jet, i.e, $0.03d$, $0.15d$, $0.30d$, $1.00d$ and $1.25d$. We see that the optical depth $\tau$, which is about 4.8 at frequency about $10^{26}\ \rm Hz$ for $z_{\rm 0}=0.03d$, reduces to $\tau<1$ for $1.25d$. Panel (b) presents $\tau$ as a function of height for different energies. We find that $\tau$ always decreases with increasing $z$. This is due to the fact that the photon field density of the stellar companion is proportional to $\propto1/z^2$, i.e., decaying with increasing height of the jet. In panel (c), we plot the optical depth as a function of orbital phase at the height $z=0.1d$ for different frequencies. The maximum of the optical depth appears at the superior conjunction, and the minimum at the inferior conjunction. The pair production of absorption interactions further promote electromagnetic cascades.

\section{Fitting Results}

\begin{figure*}[[htbp]
  \begin{center}
  \begin{tabular}{ccc}
\hspace{-0.79cm}
     \includegraphics[width=90mm,height=80mm,bb=10 200 570 579]{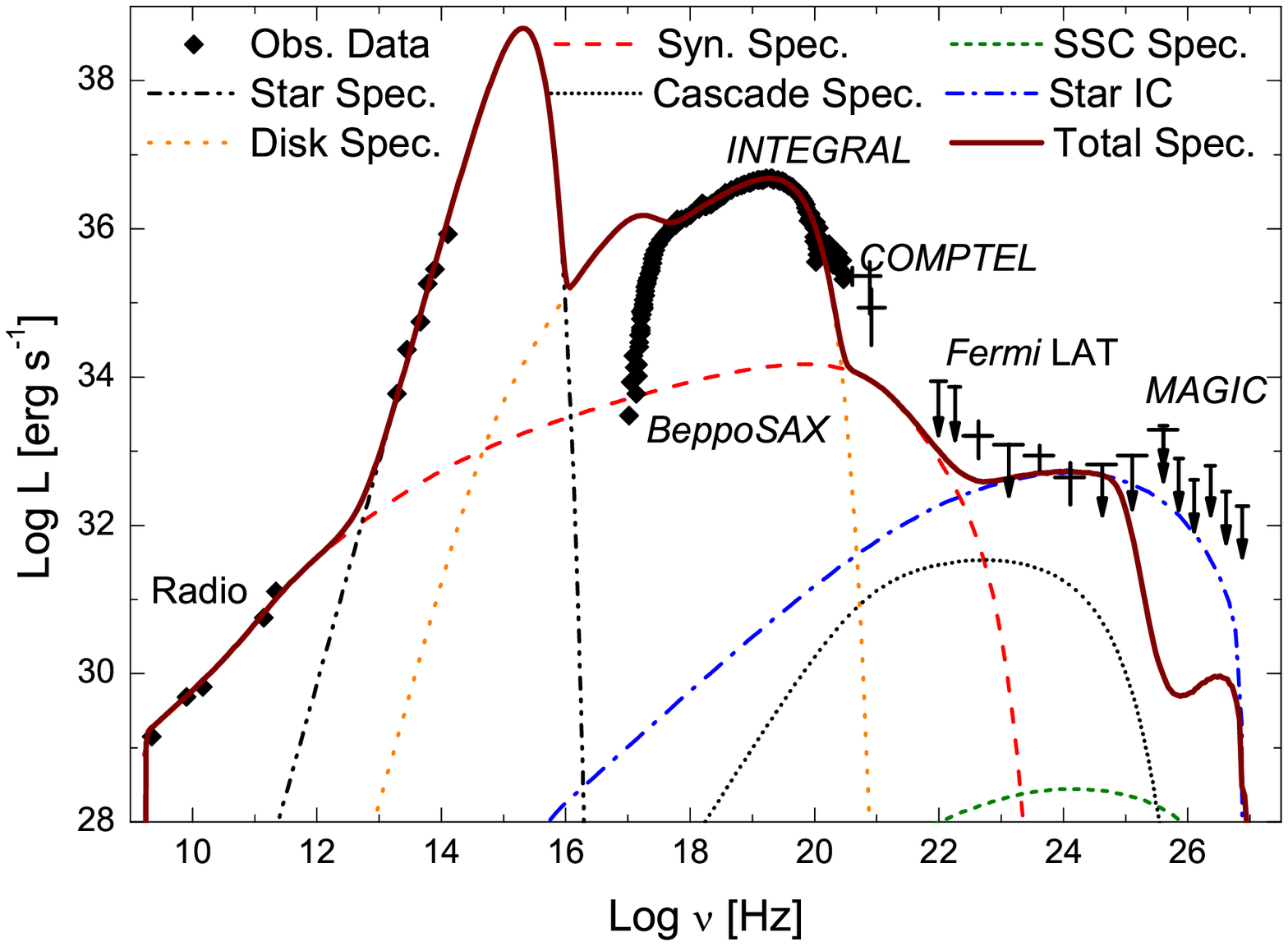} & \ \
\hspace{-0.79cm}
     \includegraphics[width=90mm,height=80mm,bb=10 200 570 579]{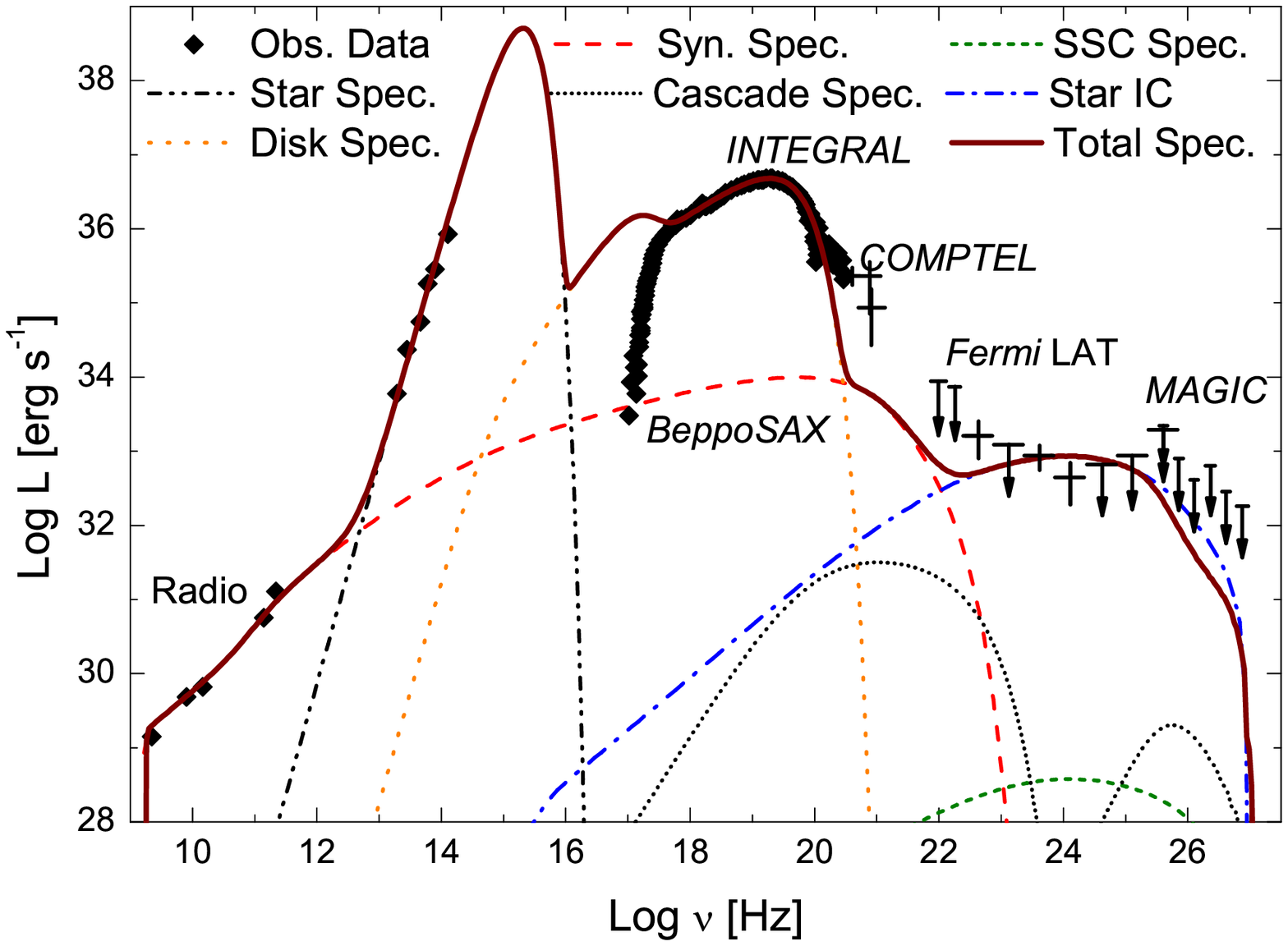}
\end{tabular}
  \end{center}
\caption{Broadband spectral energy distributions of Cygnus X--1. The parameters used in the left (right) panel are listed in Table \ref{table:cases} for Case A (Case B). The plotted observations are: radio data from \cite{Fender00}, IR fluxes from \cite{Persi80} and \cite{Mirabel96}, hard X-ray points above $5\times10^{18}\ \rm Hz$ (by \emph{INTEGRAL}) from \cite{ZAA12}, soft X-ray points below $5\times10^{18}\ \rm Hz$ (by \emph{BeppoSAX}) from \cite{DiSalvo01} (Absorption effect has been considered by an intervening medium.), soft $\gamma$-ray data points (by \emph{COMPTEL}) from \cite{McConnell02}, the \emph{Fermi} LAT measurements and upper limits from \cite{Malyshev13} and the upper limits of \emph{MAGIC} from \cite{Albert07}. The total energy spectra (thick solid line) include the first generation of electromagnetic cascade effects of $\gamma$-$\gamma$ interactions. } \label{figs:fittingmodel1}
\end{figure*}

In this section, we will employ our model to fit broadband observations ranging from radio to very high energy $\gamma$-ray bands. We want to know  which mechanism is responsible for radiation production at each band and where radiation are generated at. Therefore, we plan to carry out several possible fittings to constrain the origin of emissions. In the following fittings, the emission contribution of the first generation pairs produced in $\gamma$-$\gamma$ interaction processes is included. The related mechanisms are synchrotron emission and inverse Compton scattering of the stellar companion.

\begin{figure*}[t]
  \begin{center}
  \begin{tabular}{ccc}
\hspace{-0.89cm}
     \includegraphics[width=90mm,height=80mm,bb=10 200 570 579]{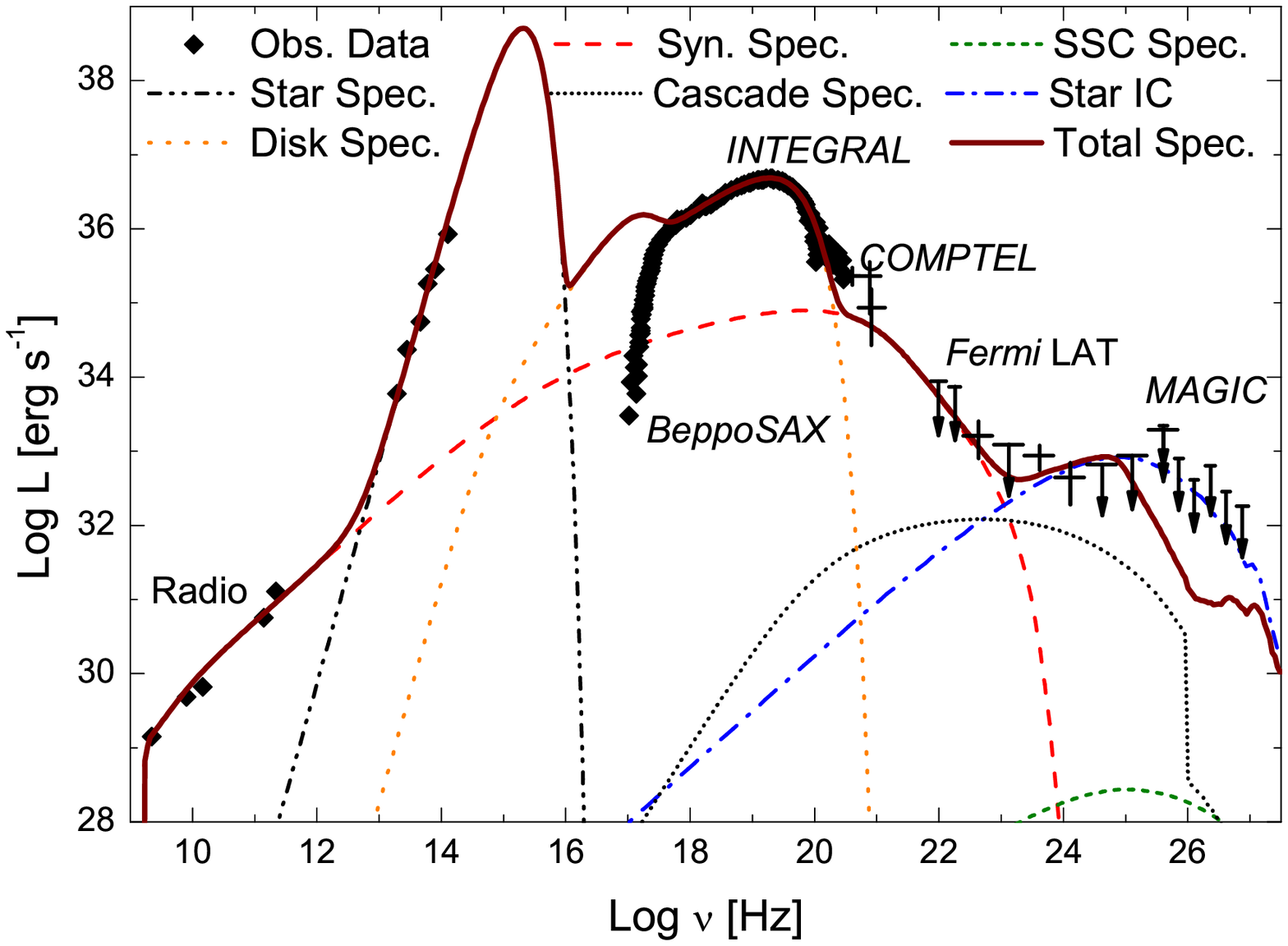} & \ \
\hspace{-0.89cm}
     \includegraphics[width=90mm,height=80mm,bb=10 200 570 579]{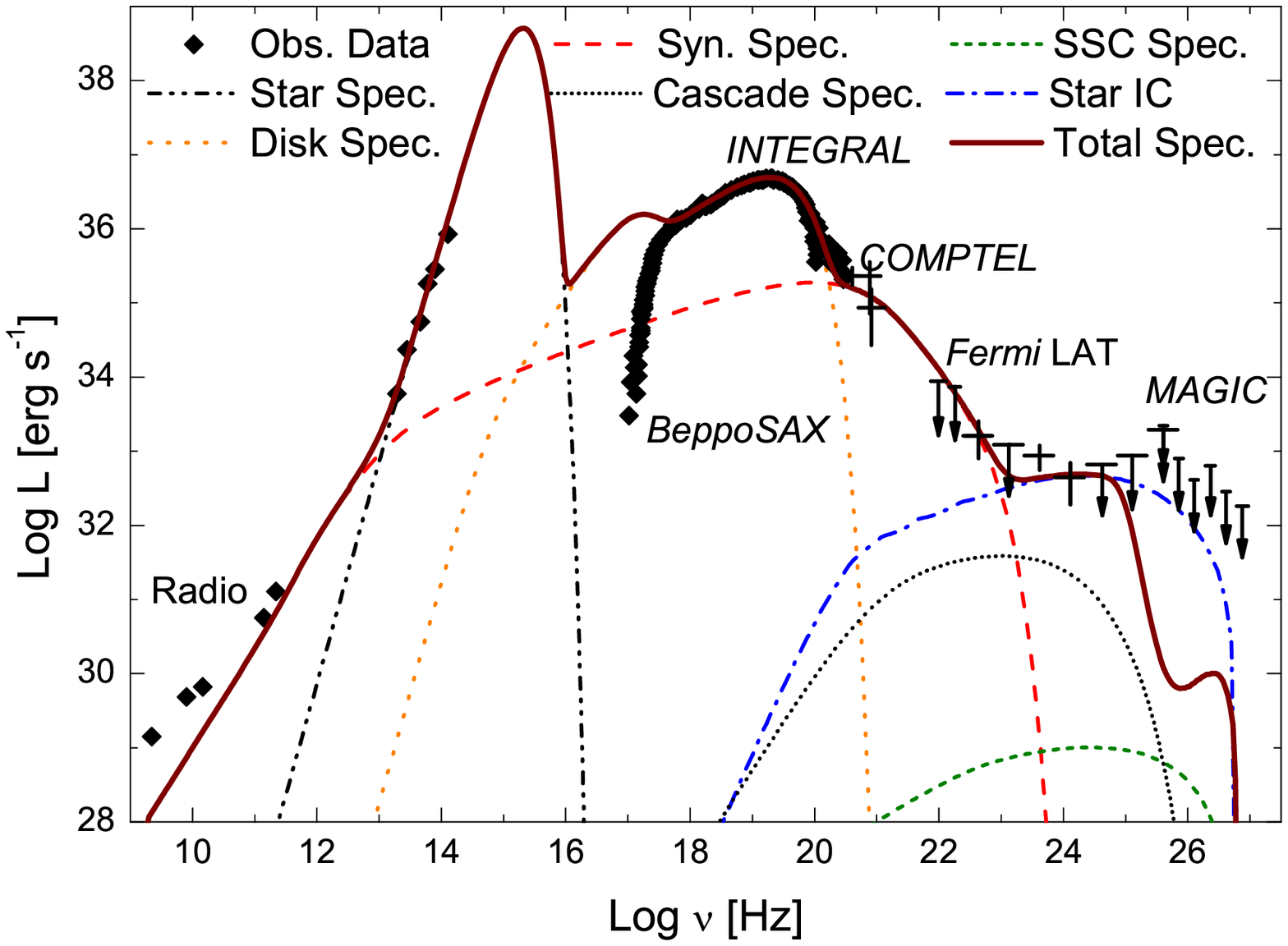}
\end{tabular}
  \end{center}
\caption{Fitting the emission spectra of Cygnus X--1. The parameters adopted in the left (right) panel are listed in Table \ref{table:cases} for Case C (Case D). Observations plotted are the same as those of Figure \ref{figs:fittingmodel1}. The total energy spectral distributions (thick solid line) include the first generation of electromagnetic cascades of $\gamma$-$\gamma$ interactions.} \label{figs:fittingmodel2}
\end{figure*}

Before presenting the fitting results, we briefly describe the fitting methods used in this work. The start location of emission region is first located. Then the magnetic field and acceleration rate are once set, we can determine an end location of electron injection by observing whether the upper limits of the maximum energy of electrons reach $\gamma_{\rm max}\sim10^7$ at what height of the jet. Furthermore, we can adjust the electron spectral index, its minimum energy and the termination of the emission region to observe the spectral shapes at radio and TeV bands by comparing to obervations. If the emission fluxes at TeV bands by the inverse Compton scattering is too high, we have to reset a higher magnetic field and repeat above steps to suppress the fluxes at TeV bands, and vice versa. The conversion efficiency of relativistic electrons $q_{\rm rel}$ is independent parameter and only affects the order of magnitude of total emission fluxes.

In Figure \ref{figs:fittingmodel1}, we present spectral energy distributions of Cygnus X--1 by comparing with observations. The fitting parameters are listed in Table \ref{table:cases} for Case A (left panel) and Case B (right panel). The plotted observational data and the upper limits from \emph{Fermi} LAT and \emph{MAGIC} telescopes are described in the caption of this figure. Observational results plotted on the all figures hold the same meaning (see also Figure \ref{figs:fittingmodel2}). Furthermore, we adopt the same display ranges for the horizontal and vertical coordinates in order to facilitate comparison. In the left panel of Figure \ref{figs:fittingmodel1}, the location of electron injection is located at $0.01d$, which is more smaller than the scale of the binary system $d$. However, in order to exclude the effects of a heavy absorption from stellar companion photons, the location of electron injection is located at $d$ for the right panel.

\begin{deluxetable}{ccccccccc}
\tabletypesize{}
%\rotate
\tablecaption{Fitting parameters for the emission spectra of Cygnus X--1.}
\tablewidth{0pt}
\tablehead{
\colhead{Case} & \colhead{$z_{\rm 0}(\rm cm)$} & \colhead{$z_{\rm inmax}(\rm cm)$} &
\colhead{$z_{\rm max}(\rm cm)$} & \colhead{$q_{\rm rel}$} & \colhead{$B'_{\rm 0}(\rm G)$} & \colhead{$p$} & \colhead{$\gamma_{\rm min}$} & \colhead{$\eta$}}
\startdata
A & $3.2\times10^{10}$  & $1.0\times10^{14}$ & $1.0\times10^{15}$  & 0.06 & $3.0\times 10^{4}$ & 1.5 & 1& $0.01$\\
B & $3.2\times10^{12}$  & $9.0\times10^{13}$ & $1.0\times10^{15}$  & 0.2 & $2.0\times 10^{2}$ & 1.6 & 1& $0.01$\\
  \tableline
C & $3.2\times10^{10}$  & $3.2\times10^{15}$ & $9.0\times10^{15}$  & 0.09 & $2.9\times 10^{4}$ & 1.5 & 1& $0.01$\\
D & $3.2\times10^{10}$  & $2.0\times10^{14}$ & $1.0\times10^{15}$  & 0.02 & $1.0\times 10^{5}$ & 1.3 & 300& $0.01$\\
\enddata
\tablenotetext{a}{Symbol indicating: $z_{\rm 0}$: initial height of electron injection, $z_{\rm inmax}$: end location of electron injection, $z_{\rm max}$: termination location of emission region, $d$: orbital radius of system, $B'_{\rm 0}$: magnetic field strength, $p$: spectral index of electron, $q_{\rm rel}$: conversion efficiency, $\gamma_{\rm min}$: minimum energy of electron, $\eta$: acceleration efficiency.}
\label{table:cases}
\end{deluxetable}

As shown in Figure \ref{figs:fittingmodel1}, radio emissions are from synchrotron emission and IR emission fluxes can be fitted with blackbody spectrum of the stellar companion. X-ray observations are fitted by using the approximated expressions given in \cite{ZAA09}, which are updated by \cite{ZAA14a} according to the latest parameters of the system. The inverse Compton scattering from the stellar companion can emit high energy and very high energy emissions, which can reproduce the observational data from \emph{Fermi} LAT and the upper limits constrained by \emph{Fermi} LAT and \emph{MAGIC} telescopes. The inverse Compton scattering of the accretion disk is negligible; it is not plotted in this figure (include also Figure \ref{figs:fittingmodel2}). We see a strong $\gamma$-$\gamma$ absorption effect in the left panel, because of the initial location of emission region being located within the binary system, in which there is a stronger photon field from the companion. Due to setting a stronger magnetic field at the base of the jet, i.e., synchrotron emission dominating radiation, which suppresses electromagnetic cascade processes. As shown in the short dotted line, cascade effects cannot contribute significant emissions to the total output spectra. Even though a weak magnetic field is set in the right panel, absorption effects due to pair production is relatively modest; thus, cascade processes have no effects to the total output spectra.

The left panel of Figure \ref{figs:fittingmodel2} presents a possible scenario that emissions at \emph{Fermi} LAT bands are produced by synchrotron emission. The fitting parameters are listed in Table \ref{table:cases} for Case C. In this regards, a slightly weak magnetic field is used comparing to the model Case A, but we have to extend electron injection heights and the termination of the emission region to more larger distances. We see that emissions of high energy pairs cannot be ignored completely in this fitting. In the above two cases, soft $\gamma$-ray (MeV tail) data from \emph{COMPTEL} campaign is not fitted.

The polarization studies of soft $\gamma$-ray emissions demonstrate that the strongly polarized signals have been observed at the 0.4 - 2\ MeV band by \cite{Laurent11}, later confirmed by \cite{Jourdain12}, which are claimed that the MeV tail emission is probably related to the optically thin synchrotron emission of high energy electrons in the jet (\citealt{Laurent11,Jourdain12}). From the theoretical point of view, there are some works to fit the MeV tail data by using the jet model (\citealt{ZAA12,Malyshev13,ZAA14b}). In particular, \cite{Russell14} claimed that the highly polarized MeV photons originate in a limb-brightened zone, at the base of the jet, with an aligned magnetic field. Very recently, \cite{Romero14} however argued that the highly polarized of the MeV tail is resulted from emissions of a corona (see also \citealt{Del13}), without need for a jet model. Alternatively, a hot accretion flow model could also reproduce the MeV tail emission flux by synchrotron emission processes (\citealt{Veledina13}). Below, we will fit the MeV tail data by using our jet model.

The fitting results for the MeV tail data are plotted in the right panel of Figure \ref{figs:fittingmodel2}. The fitting parameters are listed in Table \ref{table:cases} for Case D. In order to reproduce the MeV tail observations, we need a high value of magnetic field and a harder electron spectral index 1.3. The low energy cutoff of electron spectra are required under the limit of radio data, which are set as $\gamma_{\rm min}=300$.
Spectra of synchrotron emission and inverse Compton scattering from the stellar companion can match observations at GeV bands. Synchrotron emission fluxes can account for the MeV tail observations, which are implied by the strongly linearly polarized measurements, but the resulting spectra cannot simultaneously fit radio data. As shown in this panel, the Compton scattering of the stellar companion dominates emissions at very high energy bands, and SSC and cascade emissions are negligible.

From above four fittings to the microquasar Cygnus X--1, we know that radio observations are from synchrotron emission, optical, X-ray emissions from companion star and accretion disk, the GeV band emissions from the contributions of synchrotron emission and inverse Compton scattering of the companion, and emission fluxes at the TeV bands from inverse Compton scattering of the companion. The MeV tail observations can be reproduced by the synchrotron emission of high energy electrons provided that a high magnetic field are adopted in the model, but not simultaneously explaining radio observations.

Based on the fitting results, we know that radio emission initially takes place at the binary system scale or within it and then extend to about height $10^{15}-10^{16} \rm cm$. This is in agreement with the investigations that the large orbital modulation is observed at the radio bands (\citealt{ZAAone12}), which implies that radio emission should be emitted close to the binary system scale, and the radio structure seen by Very Large Array extends up to $z\sim10^{15}\ \rm cm$ (\citealt{Stirling01}). Owing to only having a few observational data at GeV bands and the upper limits from \emph{Fermi} LAT, it is difficult to, in this stage, firmly constrain an origin of emissions. Namely, whether these emissions are produced by synchrotron (tail) or inverse Compton scattering processes or both them. Whether they are emitted inside an orbital radius scale or outside it. A conservative inference is that the GeV band emissions should originate from the distance close to the scale of binary system, and both synchrotron and inverse Compton scattering of the stellar companion can contribute to emission outputs.

If the TeV band emissions are produced outside the binary system scale as shown in the right panel of Figure \ref{figs:fittingmodel1}, the current telescope \emph{MAGIC} or \emph{HESS} should be able to clearly detect emission signal. Although many projects about TeV band observations have been carried out, one only obtains the upper limits from \emph{MAGIC} telescope to date. We infer that emissions at TeV bands may suffer from $\gamma$-$\gamma$ absorptions from the photon field of the stellar companion, and emissions at TeV bands of electromagnetic cascades are suppressed due to existence of a strong magnetic field, which is needed to explain radio observations, as shown in the left panel of Figure \ref{figs:fittingmodel1}. Therefore, the initial height the TeV emissions are produced at, should be located within the binary system. The MeV tail emissions, producing a strong polarization signal, can be also reproduced in the model of the jet, and they should originate inside the binary system and very close to the inner region of the jet. In this case, due to the fact that the radio data cannot be simultaneously fitted, relativistic electrons producing MeV tail emission may be different from electrons that produce radio fluxes.

\section{Conclusion and Discussions}
In this work, by analogy with methods used in studies of active galactic nuclei (\citealt{ms03}), we have developed a two-dimensional, time-dependent numerical radiation model for the microquasar Cygnus X--1. By numerically solving the evolution equation for relativistic electrons, we studied spectral energy distributions of evolution electrons, taking various cooling processes into account, such as escape, adiabatic and all kinds of radiative losses. In the case of the radiative losses, we considered the synchrotron emission and its Compton scattering, external inverse Compton scattering from an accretion disk and its surrounding stellar companion. The spectral energy distributions produced at different heights of the jet are calculated, and the averaged total photon spectra are due to contributions of these individual spectra. Meanwhile, the model includes anisotropic $\gamma$-$\gamma$ absorption processes as a result of the effect of the companion photons. Furthermore, the electromagnetic cascade processes are also involved in this model.

We find that synchrotron emissions can account for the radio observations. The region of radio emission extends from the binary system scale or within this scale to the height about $z\sim10^{15}\ \rm cm$. The \emph{Fermi} LAT measurements are interpreted by the synchrotron emission and inverse Compton scattering of the companion photons. A conservative inference is that the GeV band emission should originate in the distance very close to the binary system scale. The TeV band emissions are from inverse Compton scattering processes of photons of the star; these emissions being produced within the binary system scale have suffered from an anisotropic $\gamma$-$\gamma$ absorption of the stellar photons. Furthermore, we find that emissions of positron-electron pairs are not likely to affect radiation properties of this system. But the total emission outputs could be probed by the upcoming CTA telescope. When a stronger magnetic field is used at the base of the jet, the MeV tail observation can be reproduced in our jet model by the synchrotron emission of relativistic electrons, corresponding to the emission region being located inside the binary system.

Some issues are still open, such as what fraction of the bulk kinetic energy dissipated is used to accelerate electrons, what minimum energy can an electron be effectively accelerated, what is spectral index of the accelerated electron, and what maximum energy can an electron obtain. In the current work, we used a standard first order \emph{Fermi} acceleration mechanism to constrain maximum energy of accelerated electrons. However, magnetic reconnection process or an acceleration process of the re-collimated shock produced by interactions between a jet and a stellar wind may take place.
To fit multi-band observations, we have to use a strong magnetic field strength at the base of the jet for various fitting scenarios. These values do not allow efficient electromagnetic cascades to develop. If a low enough magnetic field is used, which allows efficient electromagnetic cascade to develop, secondary pairs will produce intense radiations at the GeV bands by synchrotron emission, and at TeV bands by inverse Compton scattering of the stellar companion.

In Cygnus X-1, evidence of detection above 100 GeV of 4.1$\sigma$ significance has only been reported once right before the superior conjunction (\citealt{Albert07}). These TeV signals are around the superior conjunction of the compact object. Hence, we use the average orbital radius $d$ to calculate emission fluxes at the superior conjunction, where the absorption effect is most strong, without considering orbital modulation effects on photon spectra. Although our model are firstly developed to match the scenario in the microquasar Cygnus X--1, this model is applicable to study other microquasars provided that some corresponding modifications are conducted. This is left for future investigations.

\acknowledgments We thank the anonymous referee for his/her very constructive comments and suggestions that significantly
improved our manuscript. This work is partially supported by the National Natural Science Foundation of China (Grant Nos. 11233006 and 11363003), the Science and Technology Foundation of Guizhou Province (LKT[2012]27) and a grant S1224 of Tongren University.

\end{document}